\documentclass[twocolumn]{pasj00}
\color{black}
\begin{document}
\SetRunningHead{Murakami et al.}{Suzaku and XMM-Newton observations of the Fornax cluster}
\Received{//}
\Accepted{}

\title{
Suzaku and XMM-Newton Observations of the Fornax cluster: Temperature and Metallicity Distribution}

 \author{
   Hideyoshi \textsc{Murakami}\altaffilmark{1},
   Madoka \textsc{Komiyama}\altaffilmark{1},
   Kyoko \textsc{Matsushita}\altaffilmark{1}, 
   Ryo \textsc{Nagino}\altaffilmark{1},\\
   Takuya \textsc{Sato}\altaffilmark{1},
   Kosuke \textsc{Sato}\altaffilmark{1},
   Madoka \textsc{Kawaharada}\altaffilmark{2},
   Kazuhiro \textsc{Nakazawa}\altaffilmark{3},\\
   Takaya \textsc{Ohashi}\altaffilmark{4},
   and
   Yoh \textsc{Takei}\altaffilmark{2},
}
 \altaffiltext{1}{Department of Physics, Tokyo University of Science,
                  1-3 Kagurazaka, Shinjuku-ku, Tokyo 162-8601}
 \email{matusita@rs.kagu.tus.ac.jp}
 \altaffiltext{2}{Institute of Space and Astronautical Science, Japan Aerospace Exploration Agency,
3-1-1 Yoshinodai, Chuo-ku, Sagamihara, Kanagawa 252-5210}
 \altaffiltext{3}{Department of Physics, The University of Tokyo,
                  7-3-1 Hongo, Bunkyo-ku, Tokyo 113-0033}
 \altaffiltext{4}{Department of Physics, Tokyo Metropolitan University,
                  1-1 Minami-Osawa, Hachioji, Tokyo 192-0397}


\maketitle
\begin{abstract}
Suzaku observed a central region and five offset regions within 0.2 $r_{180}$ 
 in the Fornax cluster, a nearby poor cluster, and  XMM-Newton mapped the cluster
with 15 pointings out to 0.3  $r_{180}$.
The distributions of O, Mg, Si, S, and Fe
 in the intracluster medium (ICM) were studied with Suzaku, and
those of Fe and temperature were studied with XMM.
The temperature of the ICM gradually decreases with radius from 1.3 keV
at 0.04 $r_{180}$ to 1 keV at 0.2--0.3 $r_{180}$.
If the new solar abundances of \citet{lodd2003} and a single-temperature plasma model
are adopted,
O, Mg, Si, S, and Fe show similar abundances:
 0.4--0.6 solar within 0.02--0.2  $r_{180}$. 
This  Fe abundance is similar to those at 0.1--0.2 $r_{180}$ in
 rich clusters and other groups of galaxies.
At 0.2--0.3 $r_{180}$, the Fe abundance becomes 0.2--0.3 solar.
A two-temperature plasma model yields ICM abundances 
that are higher by  a factor of 1.2--1.5,
but gives similar abundance ratios among O, Mg, Si, S, and Fe.
The northern region has  a lower ICM temperature and higher
brightness and Fe abundance, whereas the southern region has a higher ICM
temperature and lower brightness and Fe abundance.
These results indicate that the cD galaxy may have traveled from the north
because of  recent dynamical evolution.
The cumulative oxygen- and iron-mass-to-light ratios (OMLR and IMLR)
 within 0.3 $r_{180}$ are more than an
 order of magnitude lower than those of rich clusters and
some relaxed groups of galaxies.
Past  dynamical evolution  might have hindered the strong
concentration of hot gas in the Fornax cluster's central region.
Scatter in the IMLR and similarity in the element abundances in the ICM of groups
and clusters of galaxies indicate early metal synthesis.
\end{abstract}

\section{Introduction}
\label{sec:intro}
Groups and poor clusters of galaxies
 represent the building blocks of
rich clusters and are the  best laboratories for 
the study of their thermal and chemical history, which is governed by baryons.  
An important clue to the  evolution of galaxies is the
elemental abundances  in the hot X-ray-emitting gas, e.g.,
the intracluster medium (ICM), in groups and  clusters of galaxies.
Metals in the ICM have been synthesized by supernovae (SNe) in
galaxies.  As a result, the ratios of  metal mass in the ICM to the total light
from galaxies in clusters or groups, i.e., the metal-mass-to-light ratios,
are the key parameters in investigating the chemical evolution of the
ICM\@.

Studies of the scaling relations in the clusters of galaxies have revealed strong
deviations in the  observed relations from the predictions based on
self-similar collapse \citep{Ponman99, Ponman03,
Finoguenov2007, Rasmussen2009, Johnson09, Pratt10}.
The  gas density profiles in the central regions of groups and poor
clusters are observed  to be shallower than those in the self-similar model,
and the relative entropy level is correspondingly higher than that
 in rich clusters.
These deviations are considered to be best characterized by the injection
of energy (pre-heating) into the gas before the clusters collapse
(\cite{Kaiser91}).
Based on ROSAT and ASCA data,
\citet{Ponman03} showed that groups and clusters have significant excess entropy at $r_{500}$.
\citet{Voit2003} have predicted that a smoothing of the gas density due to pre-heating
in infalling subhaloes would boost the entropy production at the accretion shock of clusters,
and an excess of entropy is generated in the cluster outskirts.
This effect due to smooth accretion should be more important for poorer systems.
However, \citet{Borgani2005} discussed that this entropy amplification effect can be
reduced by cooling.
Recently, using XMM data,  \citet{Pratt10} showed that 
at $r_{500}$, the mass dependence of the entropy excess disappeared.
\citet{Sun09} studied the entropy profiles of groups of galaxies observed with Chandra
and found that 
the difference in the entropy excess at $r_{500}$ between groups and clusters is not
as large as that by \citet{Ponman03}.
The stellar and gas mass fractions within $r_{500}$ depend on the total
system mass \citep{Vik06, Arnaud07, Sun09, Gio09}.
These studies  found that the
 stellar-to-total-mass ratios within $r_{500}$ of the groups are much 
larger than those in the clusters, whereas the gas mass fraction increases with
the system mass.

These poorer systems also differ from richer systems in that their
 iron-mass-to-light ratios (IMLR)  are systematically smaller than those in
rich clusters (\cite{Makishima2001}).
The metal distribution in the ICM is 
a tracer of the history of  gas heating,
because both metal enrichment and heating timescales
determine the metal distribution in the ICM.
The Chandra and XMM observations of nearby groups of galaxies with cool cores found that
the Fe abundances of the groups declines with the radius
 and metal-mass-to-light ratios of groups
are much smaller than those of the clusters
\citep{Rasmussen2007, Finoguenov2007, Rasmussen2009,Johnson11}.
On the basis of the Chandra data,
\citet{Rasmussen2009} discussed the effect of feedback and 
the history of the 
more extended star formations in less massive systems. 
With XMM observations, \citet{Johnson11} found a difference in the abundance
profiles of the cool core and non-cool core groups and discussed 
the effect of  mixing driven by active galactic nuclei (AGN)
 within the central regions.

O and Mg are predominantly synthesized   in SN II, whereas Fe and Si are
synthesized in both SN Ia and SN II\@. Therefore, abundance measurements
spanning the range of species from O to Fe are  required for the
unambiguous determination of the formation history of massive stars.
XMM-Newton provided a means of  constraining the O and Mg abundances of some
systems \citep{Matsushita2003,Tamura2003,Matsushita2007b,Aurora2009}.
However, reliable results have been obtained
only for the central regions of very bright clusters or groups of
galaxies dominated by cD galaxies.
 The X-ray imaging spectrometer (XIS; 
 \cite{Koyama2007}) onboard Suzaku \citep{Mitsuda2007} offers an improved
line spread function because of its very small low-pulse-height tail in the
energy range below 1~keV coupled with a very low background.
Therefore, especially for regions of low surface brightness or
equivalent width,  XIS provides better sensitivity to O lines.
The instrumental Al line of the MOS detectors
 on XMM-Newton causes problems in
 measuring the Mg abundance in somewhat fainter systems.

The oxygen-mass-to-light ratios (OMLRs) as
well as the IMLRs of several clusters of galaxies and several groups of
galaxies out to 0.2--0.3~ $r_{180}$ were measured  with Suzaku satellite
\citep{Matsushita2007a, Tokoi2008, kSato2007a,kSato2008, kSato2009a, kSato2009b,
Komiyama2009, kSato2010}.  
The difference in the OMLRs  between
groups and clusters is a factor of about 3--6, and tends to be 
larger than the IMLR difference, which is a factor of 2--3 (\cite{Komiyama2009}).

The Fornax cluster is a nearby poor cluster with an ICM temperature of
1.3--1.5keV (\cite{Scharf2005}). The X-ray emission shows an
asymmetric spatial distribution, and the cD galaxy, NGC 1399, is offset
from the center \citep{Pao2002, Scharf2005}, which may be
related to  large-scale dynamical evolution such as infall motions of
galaxies into the cluster (\cite{Dunn2006}). The Chandra observations
suggest that  relative motion may occur between NGC 1399 and the ICM,
and that the second brightest elliptical galaxy, NGC 1404, is moving
supersonically in the ICM (\cite{Scharf2005}; \cite{Machacek2005}). 
The Fe and Si abundances of the ICM
 within $\sim 50 \,\mathrm{kpc}$ of NGC 1399 were measured with XMM-Newton (Buote
2002). 
The OMLR and IMLR within 0.13 $r_{180}$ derived from 
 early Suzaku observations of two fields of the Fornax cluster (\cite{Matsushita2007a})
are the smallest among those in the groups of galaxies observed with Suzaku.

In this paper, we describe our study of the ICM of the Fornax cluster for 
 regions  within 0.035--0.2 $r_{180}$ observed with Suzaku
and  the temperature and Fe abundance out to 0.3 $r_{180}$ observed with XMM-Newton.
In Section 2, we summarize the observations and data preparation.
Section 3 describes our analysis of the data, and in Section 4, the temperature
and the  O, Mg, Si, S, and Fe abundances are determined.
We discuss our results in Section 5.

We use the Hubble constant $H_{0} = 70~\mathrm{km~s^{-1}~Mpc^{-1}}$. 
The distance to the Fornax cluster is $D_\mathrm{L} = 19.8$~Mpc and $1'$
corresponds to 5.70~kpc. The virial radius,
$r_{180}=1.95~h_{100}^{-1} \sqrt{k \langle T \rangle /10~\mathrm{keV}}$~Mpc
 (\cite{Markevitch1998}; \cite{Evrard1996}), is about
1~Mpc for the average temperature $k \langle T \rangle = 1.3$~keV.  
In this paper, we
 use the new abundance table from \citet{lodd2003}.
  The  O and Fe abundances are about 1.7 times
and 1.6 times higher than those of  \citet{angr1989}, respectively.
Unless otherwise specified, errors are quoted at 90\% confidence for the
single parameter of interest.

\section{Observations}
\label{sec:fornax_obs}
\subsection{Suzaku observations}
\begin{figure*}
\centerline{
    \FigureFile(48mm,70mm){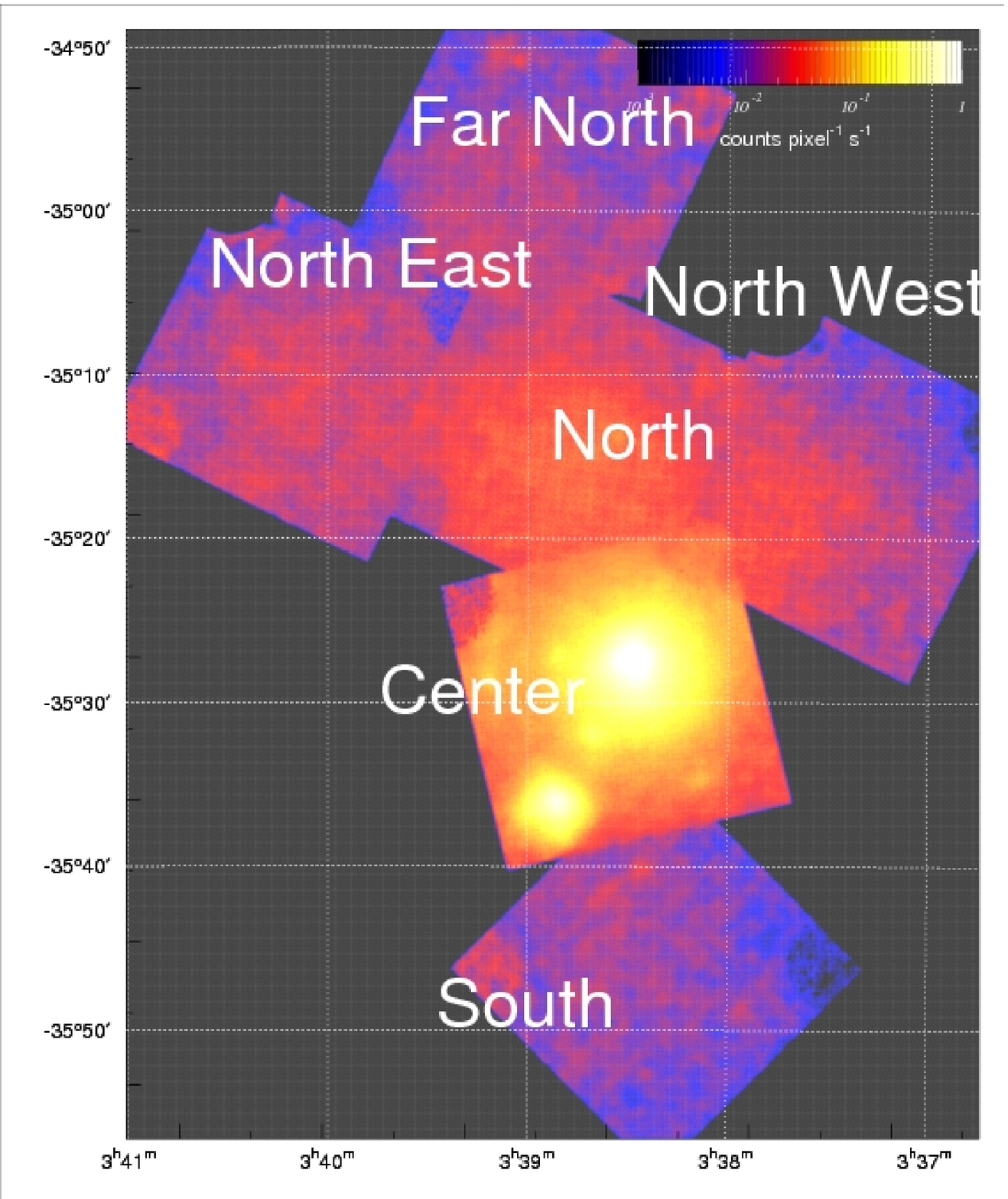}
    \FigureFile(70mm,75mm){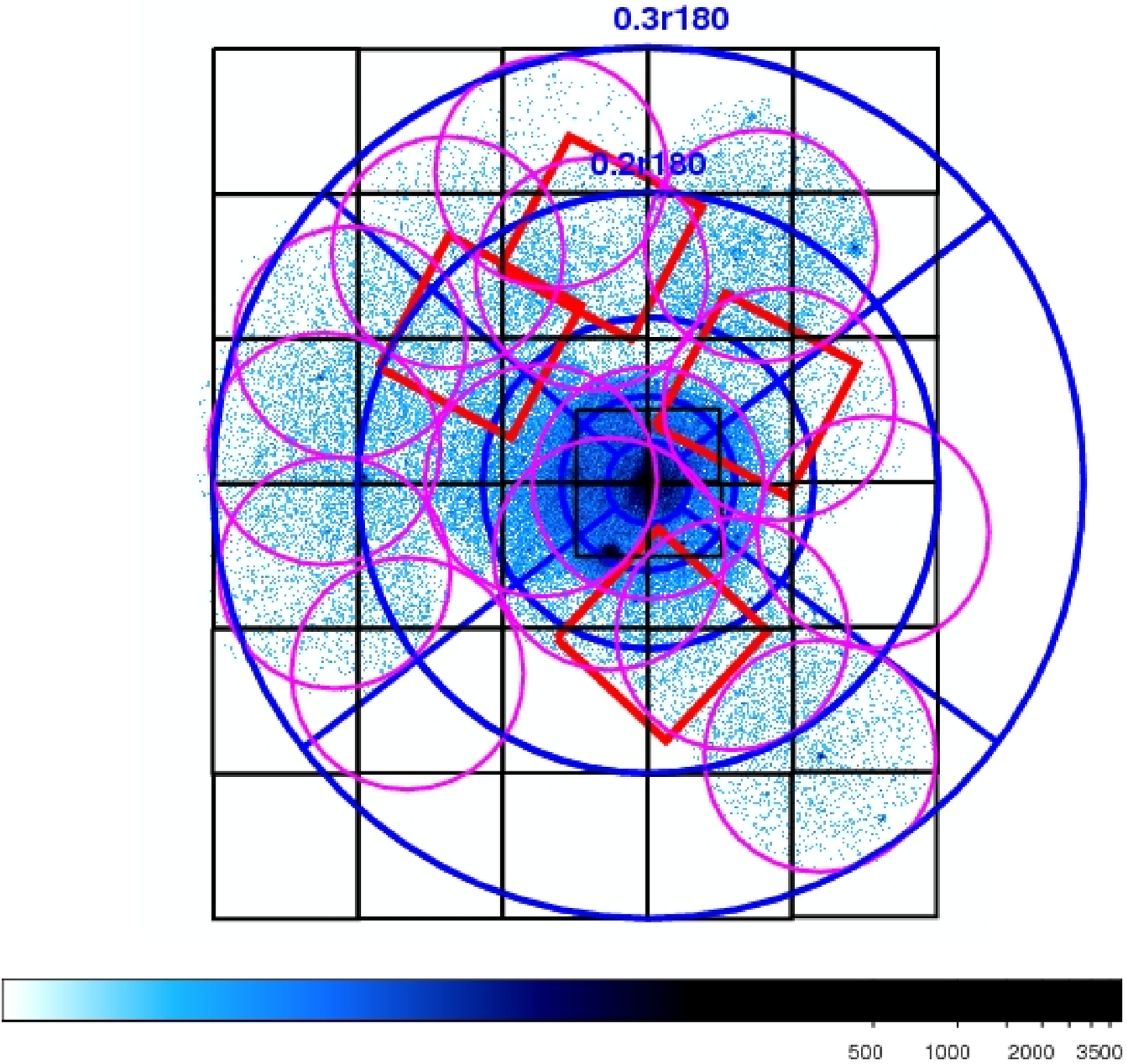}
    \FigureFile(55mm,75mm){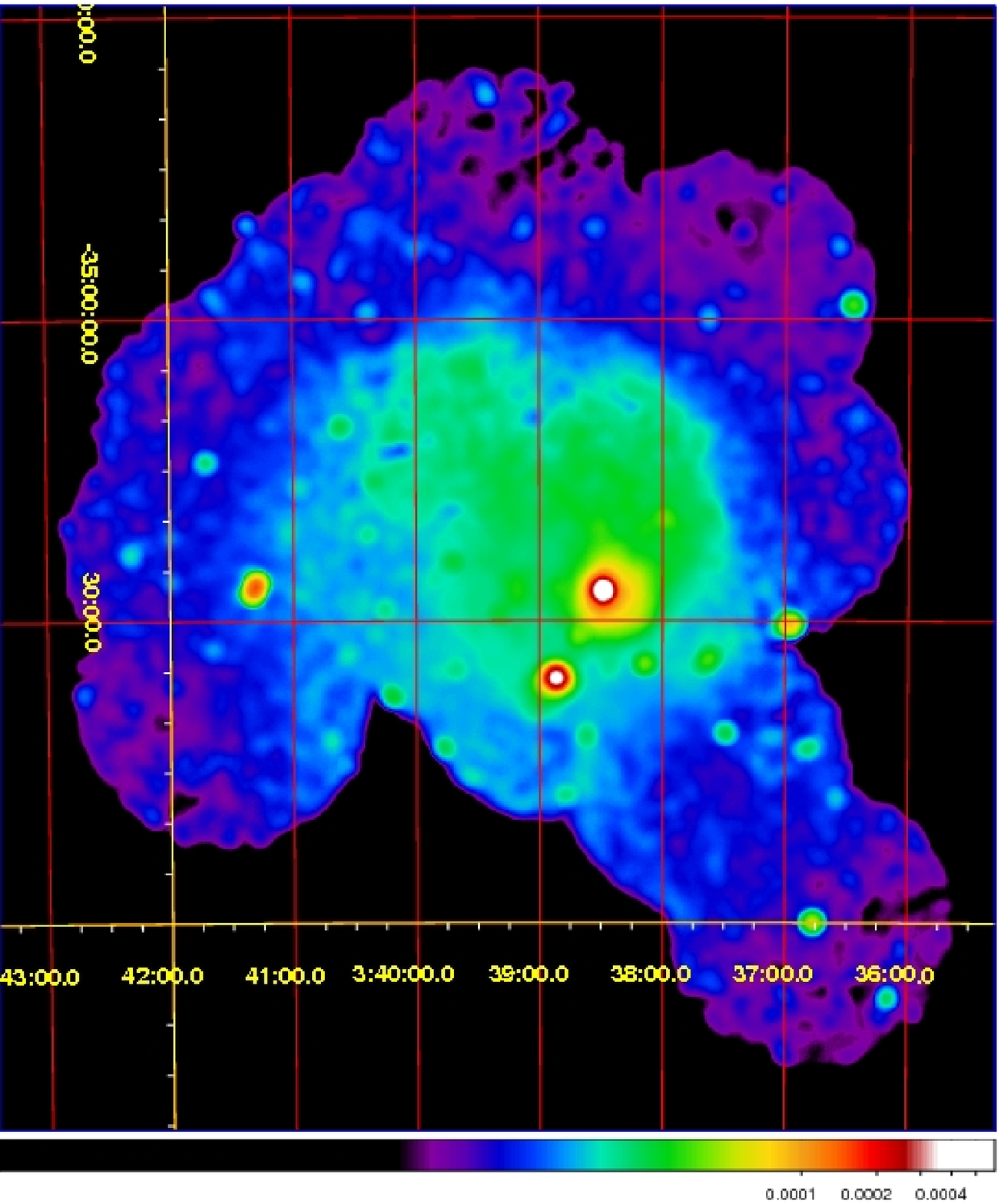}
 }
 \caption{
(left)
Suzaku-XIS (0.5--4.0~keV) image of the Fornax cluster. 
 The NXB was subtracted
 and the difference in  exposure times was corrected.
 The Cosmic X-ray background (CXB) was not subtracted.
(middle)
Raw XMM-MOS image (0.8--1.2~keV) of the Fornax cluster.
Magenta  circles correspond to field of view of the XMM observations.
Red squares indicate the four offset observations with Suzaku.
Blue pie and black square regions summarize the accumulation area of 
spectral analysis.
(right)
Exposure- and vignetting-corrected and 
adaptively smoothed XMM-MOS image (0.8--1.2~keV).
The NXB and CXB were not subtracted.
}
\label{fig:fornax_mosaic}
\end{figure*}

Suzaku performed six pointing observations of the Fornax cluster, as
summarized in Table 
\ref{tab:fornax_obsinfo}.
The first observation (hereafter, Center field) was carried out on 
2005 September with the pointing direction 2$'$ south and 1$'$ east of NGC 1399.
The second one (hereafter, North field) was centered 13$'$ north and 4$'$ east of NGC 1399, and was carried out on 2006 January.
Four additional observations were centered $\sim 30'$ north (hereafter, Far North field),
and $\sim$ 17$'$--27$'$ south, northwest, and northeast (hereafter,
South, North West, and North East fields, respectively) of NGC 1399.
The left panel of Figure \ref{fig:fornax_mosaic} shows a 0.5--$4.0$ keV image 
obtained with these Suzaku observations.
The observed region covers a distance of about about $42'$,
or $\sim 240~{\rm kpc}$ toward the north from NGC 1399:
this distance corresponds to 0.24 $r_{180}$.
To constrain emissions from our Galaxy, we also observed
two  fields offset of $\sim 5$ degrees  from NGC 1399
(hereafter, Galactic1 and Galactic2 fields)
in 2007 June. The observation log  is also shown in Table
\ref{tab:fornax_obsinfo}.

The  XIS was operated in the nominal mode during these observations. 
We included the data formats of both $5\times5$ and
$3\times3$ editing modes in our analysis using {\tt
XSELECT(Ver.~2.4a)}.  The analysis was performed using {\tt
HEAsoft(Ver.~6.6.3)} and {\tt XSPEC(Ver.~11.3.2ag)}.  After 
the standard data selection criteria are applied,
 the exposure times of the four offset fields are 35--56 ks.
Those of the Galactic1 and Galactic2 fields are $\sim$ 20 ks.

\begin{table*}
\begin{center}
\caption{
Suzaku and XMM observations of the Fornax cluster and background fields
}
\label{tab:fornax_obsinfo}
\begin{tabular}{lllll}
\hline              
Fields & Seq. No. & (R.A., Dec.) in J2000.0& Date of obs. & Exp. time (after screenings) \\
  \hline
\multicolumn{5}{l}{Suzaku observations of the Fornax cluster}\\
Center
 & 100020010
 & (\timeform{3h38m35s.7},~\timeform{-35D28'14''.5})
 & 2005/09/13 & 76 ks \\
North 
 & 800002010
 & (\timeform{3h38m51s.9},~\timeform{-35D14'23''.6})
 & 2006/01/05 & 78 ks \\
Far North 
 & 802021010
 & (\timeform{3h38m55s.9},~\timeform{-34D57'32''.4})
 & 2008/01/14 & 56 ks \\
South 
 & 803006010
 & (\timeform{3h38m19s.6},~\timeform{-35D45'49''.1})
 & 2008/07/15 & 35 ks \\
North West 
 & 803007010
 & (\timeform{3h37m25s.7},~\timeform{-35D16'52''.7})
 & 2008/07/16 & 41 ks \\
North East
 & 803008010
 & (\timeform{3h40m07s.5},~\timeform{-35D09'04''.1})
 & 2008/07/17 & 41 ks \\
\hline
\multicolumn{5}{l}{Suzaku observations of the Fornax Galactic field}\\
Galactic1
 & 802037010
 & (\timeform{3h13m11s.0},~\timeform{-37D40'48''.0})
 & 2007/06/28 & 20 ks \\
Galactic2
 & 802040010
 & (\timeform{3h19m57s.7},~\timeform{-32D04'18''.8})
 & 2007/06/29 & 21 ks \\
\hline
\multicolumn{4}{l}{XMM observations of the Fornax cluster} &MOS1, MOS2, PN\\
A
 & 0550930101
 & (3$^\mathrm{h}$39$^\mathrm{m}$02$^\mathrm{s}$.4,~-35$^{\circ}$01$'$55$''$.2)
 & 2008/06/28
 & 10.6, 11.2, 7.0 ks \\
B
 & 0550930201
 & (3$^\mathrm{h}$39$^\mathrm{m}$26$^\mathrm{s}$.2,~-34$^{\circ}$49$'$37$''$.2)
 & 2008/06/27
 & 7.6, 6.7, 3.5 ks \\
C
 & 0550930301
 & (3$^\mathrm{h}$40$^\mathrm{m}$27$^\mathrm{s}$.1,~-34$^{\circ}$59$'$16$''$.8)
 & 2008/07/17
 & 11.4, 11.6, 8.0 ks \\
D
 & 0550930401
 & (3$^\mathrm{h}$41$^\mathrm{m}$25$^\mathrm{s}$.0,~-35$^{\circ}$10$'$04$''$.8)
 & 2009/02/09
 & 14.5, 15.1, 11.7 ks \\
E
 & 0550930501
 & (3$^\mathrm{h}$41$^\mathrm{m}$40$^\mathrm{s}$.8,~-35$^{\circ}$22$'$51$''$.6)
 & 2009/02/23
 & 18.1, 17.8, 14.6  ks \\
F
 & 0550930601
 & (3$^\mathrm{h}$41$^\mathrm{m}$35$^\mathrm{s}$.0,~-35$^{\circ}$37$'$48$''$.0)
 & 2009/02/24
 & 17.7, 17.9, 13.7 ks \\
G
 & 0550930701
 & (3$^\mathrm{h}$40$^\mathrm{m}$52$^\mathrm{s}$.1,~-35$^{\circ}$50$'$02$''$.4)
 & 2009/02/24
 & 0, 0, 0 ks \\
J
 & 0550931001
 & (3$^\mathrm{h}$37$^\mathrm{m}$38$^\mathrm{s}$.9,~-35$^{\circ}$45$'$18$''$.0)
 & 2008/06/25
 &  19.1, 19.4, 11.7ks \\
L
 & 0550931201
 & (3$^\mathrm{h}$36$^\mathrm{m}$15$^\mathrm{s}$.6,~-35$^{\circ}$32$'$56$''$.4)
 & 2008/06/25
 & 0, 0, 0 ks \\
N
 & 0550931401
 & (3$^\mathrm{h}$37$^\mathrm{m}$11$^\mathrm{s}$.5,~-35$^{\circ}$17$'$34$''$.8)
 & 2008/06/26
 & 11.0, 11.4, 8.0  ks \\
NGC 1399
 & 0400620101
 & (3$^\mathrm{h}$38$^\mathrm{m}$29$^\mathrm{s}$.1,~-35$^{\circ}$27$'$03$''$.0)
 & 2006/08/23
 & 99.6, 102.7, 53.2 ks \\
NGC 1404
 & 0304940101
 & (3$^\mathrm{h}$38$^\mathrm{m}$51$^\mathrm{s}$.9,~-35$^{\circ}$35$'$39$''$.8)
 & 2005/07/30
 & 24.6, 15.0, 17.0  ks \\
LP 944-20
 & 0055140101
 & (3$^\mathrm{h}$39$^\mathrm{m}$34$^\mathrm{s}$.60,~-35$^{\circ}$25$'$51$''$.0)
 & 2001/01/07
 & 43.0, 43.2, 36.5 ks \\
RXJ 0337-3457
 & 0210480101
 & (3$^\mathrm{h}$37$^\mathrm{m}$24$^\mathrm{s}$.70,~-34$^{\circ}$57$'$29$''$.0)
 & 2005/01/04
 & 44.3, 44.4, 37.9 ks \\
NGC 1386
 & 0140950201
 & (3$^\mathrm{h}$36$^\mathrm{m}$45$^\mathrm{s}$.4,~-35$^{\circ}$59$'$57$''$.0)
 & 2002/12/29
 & 15.9, 15.9, 12.8 ks \\
\hline
\end{tabular}
\end{center}
\end{table*}

The spectra of the Center and North fields were accumulated within concentric rings,
6$'$--13$'$ and 13$'$--26$'$, centered on NGC 1399.
The spectra  of the Far North, South, North West and North East fields
 were  accumulated over the field of view of XIS.
Each spectrum was binned 
 to  observe details in metal lines,
and each spectral  bin contained 50 or more counts.

The response of the X-ray telescope (XRT) and XIS for each spectrum was calculated using the
 {\tt xisrmfgen} response matrix file (RMF) generator,  version
{\tt 2009-02-18}. 
The ancillary response files (ARF) were calculated using
  {\tt xissimarfgen} \citep{Ishisaki2007}, version {\tt 2009-01-08}, assuming flat
 emission, because the Fornax cluster is much more extended than the field
 of view of the XIS,
and ARFs assuming flat-sky emission, a $\beta-$model profile, and a point-source
are almost the same within the energy range of 0.4 to 5 keV, except for normalization.
Slight degradation of the energy resolution was
considered in the RMF, and decrease in the low-energy transmission of
the XIS optical blocking filter (OBF) was included in the ARF\@. 

The non-X-ray background (NXB) was subtracted from the spectra using
a database of  night Earth observations \citep{Tawa2008}.
  We used the spectra within an energy range of 0.4--5.0$~{\rm
keV}$, because above $5.0~{\rm keV}$, background components dominate 
the X-ray emission.

\subsection{XMM-Newton observations}

We analyzed 15 pointing XMM-Newton observations of the Fornax cluster
(Table \ref{tab:fornax_obsinfo}).

The middle and right panels of Figure  \ref{fig:fornax_mosaic} show a 0.8--1.2 keV  MOS image.
The NXB and Cosmic X-ray background (CXB) 
were not subtracted because this energy band is dominated by
the ICM emission.
The observed region covers out to 0.3 $r_{180}$ in the east, north and southwest
directions.
We used the PN, MOS1, and MOS2 detectors and
 selected events with patterns smaller than 5 and 13 for the PN 
and MOS, respectively.
To screen  background flares, we constructed  count rate histograms of
PN and MOS.
Then, we  fitted  each histogram with a Gaussian and selected the time 
within 2.5 $\sigma$ of the mean for each histogram.
The total exposures after background flares are summarized in 
Table \ref{tab:fornax_obsinfo}.
Because the observations of  fields G and L were completely dominated by very high background flares, no exposure time remained.

Spectra were accumulated in  pie regions of north,
east, south, and west,  centered on NGC 1399 as summarized in Figure   \ref{fig:fornax_mosaic}.
We also accumulated spectra in the square regions of
0.1 $r_{180}\times$ 0.1 $r_{180}$ (17.5$'\times$17.5$'$).
Each square region surrounding NGC 1399 is divided into 
two: one is a small square region  0.05 $r_{180}\times$ 0.05 $r_{180}$
and the other, as shown in Figure \ref{fig:fornax_mosaic}.
When accumulating spectra, luminous point sources and  NGC 1404 were excluded.
Although we have not screened hot chips of MOS \citep{Kuntz2008}, we
have verified that exclusions of the hot chips do not affect any results.
RMF and ARF corresponding to 
each spectrum were calculated using SAS v8.0.0.
The spectral analysis also used the XSPEC$\_$v11.3.2ag package.

\section{Spectral Analysis}
\label{sec:ana}

\subsection{Galactic fields observed with Suzaku}
\label{sec:ana_Galactic}
To estimate the Galactic emission, we first analyzed the two Galactic fields
observed with Suzaku.
The Galactic emission, which includes the local hot bubble,
the Milky Way halo, and solar wind charge exchange,
 is empirically fitted with a two-temperature
plasma model with redshift $=0$ (\cite{Lumb2002}; \cite{Yoshino2009}).
Therefore, we used the two-temperature APEC
thermal model  for the  Galactic emission, with a power-law model
for the CXB, and fitted the observed spectra
from the Galactic1 and Galactic2 fields.
The temperature and normalization of the two components
in the Galactic emission were left free, with the metal abundance fixed to
the solar level.
CXB was modeled by a power-law spectrum with a photon index $\Gamma = 1.4$.
The normalization of the power-law was allowed to be free.
The model spectra, except for the lower-temperature APEC component, 
were subjected to a common interstellar absorption $N_{\rm H}$, 
fixed at the Galactic value in the direction of  each field  by \citet{Dickey}.

The results of the spectral fits are shown in Table
\ref{tab:fornax_Galkt}.
Except for the instrumental Al line at 1.5 keV,
this model reproduced the spectra well.
The derived temperatures of the two APEC components and the
ratio of the normalizations
of the two Galactic fields are consistent with each other.
The derived temperatures,  $\sim$ 0.1 keV and $\sim$ 0.3 keV,
are  consistent with the typical values for the Galactic emission.

\begin{table}
\begin{center}
\caption{
The results of the spectral fitting of the Galactic emission with Suzaku.~~~~~~~~~~~~~  
}
\label{tab:fornax_Galkt}
\begin{tabular}{lcccc}
\hline
  Region
 & $kT_{\rm 1}$
 & $kT_{\rm 2}$
 & Ratio of
 & $\chi^2$/d.o.f. \\
 & (keV) & (keV)
 & Norm\footnotemark[$\ast$]
 & - \\ \hline
  Galactic1
 & $0.11^{+0.02}_{-0.02}$
 & $0.37^{+0.09}_{-0.07}$
 & $0.11^{+0.09}_{-0.06}$
 & $151/129$\\
   Galactic2
 & $0.09^{+0.02}_{-0.02}$
 & $0.28^{+0.05}_{-0.04}$
 & $0.10^{+0.15}_{-0.06}$
 & $140/113$\\
\hline
\multicolumn{5}{l}{\hbox to 0pt{\parbox{85mm}{\footnotesize 
\par\noindent
\footnotemark[$\ast$]
The ratio of normalizations of higher and lower
temperature components.
}\hss}}
\end{tabular}
\end{center}
\end{table}

\subsection{The Fornax cluster observed with Suzaku}

\begin{figure*}
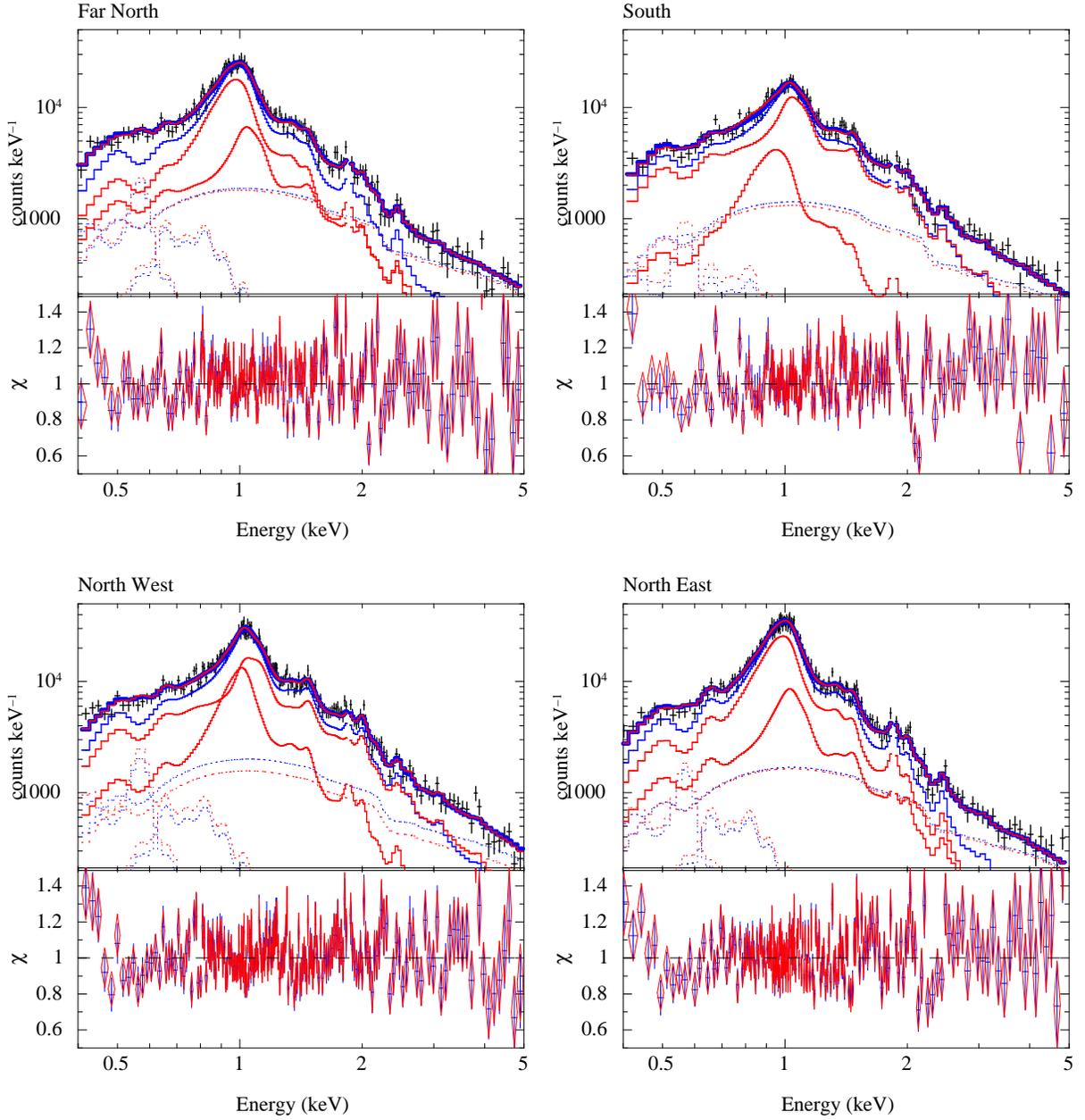

  \begin{center}
   \FigureFile(80mm,50mm){PASJ4026_figure2a.ps}
   \FigureFile(80mm,50mm){PASJ4026_figure2b.ps}
   \FigureFile(80mm,50mm){PASJ4026_figure2c.ps}
   \FigureFile(80mm,50mm){PASJ4026_figure2d.ps}
  \end{center}
\caption{
Spectra of the four offset fields from the XIS-1 instruments fitted with
the 1T (blue) or 2T (red) model for the ICM (solid lines)  and
background components (dashed lines) including the power-law
and the Galactic components.
Lower panels show the data-to-model ratios for the 1T (blue crosses) and
the 2T (red diamonds) model fits.
}
\label{fig:fornax_1Tspec}
\end{figure*}

We first assumed an ICM in each region consisting of a single-temperature vAPEC
\citep{Smith2001} model (hereafter, the 1T model).  
We fitted all the XIS spectra of each region
with  the 1T  model for the ICM,  
a power-law model for the CXB, and the two-temperature APEC model
for the Galactic components.
The temperature, abundances, and normalization of the ICM component
of each field were allowed to vary.
Here, the metal abundances
of He, C, and  N were fixed to the solar values. We divided
the other metals into six groups:
 O; Ne; Mg and Al; Si; S, Ar, and Ca; Fe and Ni.
The spectral components except for the lower-temperature Galactic emission were subjected to a common
interstellar absorption, $N_{\rm{H}}$, which was allowed to vary for each instrument,
considering the systematic uncertainties of  contaminants on the XIS detectors.
The temperatures of the two Galactic components were fixed at 0.1 keV and 0.3 keV, respectively, 
and the ratio of the normalizations of the higher- and lower-temperature components
was fixed at 0.1, which was the best-fit value for the two Galactic fields.

The results are shown in  Table
\ref{tab:fornax_ktno} and the fitted spectra  of the four 
new offset fields are shown in Figure \ref{fig:fornax_1Tspec}.
The derived reduced $\chi^2$ values were $\sim 1$, and
the spectra of the Fornax cluster were
consistently reproduced by the sum of the ICM model and this Galactic component.

We also applied a two-temperature model for the ICM (hereafter, the 2T model),
where the abundances of each metal in the two components were assumed to have the same value.
The results are  shown in Table \ref{tab:fornax_ktno} and Figure \ref{fig:fornax_1Tspec}.
Compared to the 1T model,
the reduced $\chi^2$ decreased slightly by  several percent, 
whereas, the metal abundances  increased by several tens of percent.
As shown in Table 3, F-test probabilities
 support an addition of the second temperature component in most of the systems.
However,
considering the possible systematic uncertainties in the Fe-L and background
components, we cannot determine whether the 2T model is better than the 1T model
because the 1T model  still represents the spectra fairly well.

To study the systematic uncertainties in  the modeling of the Galactic components,
we fitted the spectra with the 1T or 2T model for the ICM in  a similar manner,
except that the normalizations of the two Galactic components of each field were allowed to be free.
The temperature and abundances of Ne, Mg, Si, S, and Fe,
and the weighted averages of O/Fe, Ne/Fe, Mg/Fe, Si/Fe, and S/Fe  were almost the same
within 10--20\%,
although the best-fit values of the  O abundances were changed by 0.1--0.2 solar.

\begin{table*}
\begin{center}
\caption{
The ICM temperature, $\chi^2$, and elemental abundances 
derived from the 
spectral fits of Suzaku for the Fornax cluster.
}
\label{tab:fornax_ktno}
\begin{tabular}{lccccclcc}
\hline
Field 
 & $r$ & model 
 & $kT$ & $kT$
 & $\chi^2$/d.o.f. &  
\multicolumn{2}{l}{
F-test probability$^\ast$}
 \\
 & (arcmin) & 
 &  (keV)  & (keV) &  & 
 \\ \hline
Center & 6--13 & 1T &   $1.32^{+0.01}_{-0.01}$ &  & 1065/934\\
Center & 6--13 & 2T & $0.81^{+0.03}_{-0.04}$ & $1.50^{+0.05}_{-0.06}$   & 979.4/933 & 1$\times 10^{-18}$\\
North  & 6--13 & 1T &   $1.27^{+0.01}_{-0.01}$ &  & 754.4/734\\
North  & 6--13 & 2T & $0.84^{+0.22}_{-0.19}$ & $1.32^{+0.23}_{-0.05}$   & 731.3/732& 1$\times 10^{-5}$\\
North  & 13--26 & 1T &  $1.16^{+0.01}_{-0.01}$ &  & 1058/956\\
North  & 13--26 & 2T & $1.02^{+0.03}_{-0.11}$ & $1.45^{+0.16}_{-0.24}$ &   975.4/954 & 1$\times 10^{-17}$\\
North West 
 & 7--28
  & 1T 
 & $1.33^{+0.01}_{-0.01}$
& & 521.1/428 &
 &  
 \\
North West 
 & 7--28
  & 2T 
&  $1.09^{+0.16}_{-0.08}$
&  $1.72^{+0.19}_{-0.16}$
& 504.0/427 & 0.0002\\
South 
 & 10--31
& 1T 
 & $1.27^{+0.02}_{-0.02}$
& & 325.7/275
 &  &  
 \\
South 
 & 10--31
& 2T &$0.84^{+0.15}_{-0.05}$ &$1.48^{+0.17}_{-0.14}$ & 307.0/274 & 6$\times 10^{-5}$\\
North East
 & 17--36
& 1T 
 & $1.04^{+0.01}_{-0.01}$
& & 403.1/384
 &  &  
 \\
North East 
 & 17--36
& 2T &  $1.00^{+0.04}_{-0.13}$ &  $1.29^{+1.08}_{-0.77}$
& 395.3/383 & 0.006\\
Far North
 & 20--42
 & 1T
& $1.05^{+0.01}_{-0.01}$
 & & 382.6/340
   &  & 
 \\
Far North
 & 20--42
 & 2T & $0.98^{+0.05}_{-0.13}$ & $1.43^{+0.72}_{-0.22}$ & 365.9/339 & 0.0001\\
\hline
Field 
 & $r$ & model 
 & O & Ne & Mg & Si & S& Fe
 \\
 & (arcmin) &
 & (solar) & (solar) & (solar) & (solar)
 & (solar) & (solar)
 \\ \hline
Center & 6--13 & 1T & $0.30^{+0.07}_{-0.07}$ & $0.10^{+0.14}_{-0.10}$ & $0.25^{+0.06}_{-0.06}$ & $0.33^{+0.04}_{-0.04}$ & $0.38^{+0.05}_{-0.05}$ & $0.38^{+0.02}_{-0.02}$ \\
Center  & 6--13 & 2T & $0.41^{+0.09}_{-0.09}$ & $0.75^{+0.25}_{-0.23}$ & $0.45^{+0.09}_{-0.09}$ & $0.46^{+0.06}_{-0.06}$ & $0.47^{+0.07}_{-0.07}$ & $0.54^{+0.05}_{-0.05}$ \\
North & 6--13 & 1T & $0.44^{+0.12}_{-0.11}$ & $0.31^{+0.19}_{-0.19}$ & $0.34^{+0.07}_{-0.07}$ & $0.45^{+0.05}_{-0.05}$ & $0.52^{+0.07}_{-0.07}$ & $0.53^{+0.03}_{-0.03}$ \\
North & 6--13 & 2T & $0.54^{+0.14}_{-0.13}$ & $0.78^{+0.37}_{-0.29}$ & $0.47^{+0.11}_{-0.10}$ & $0.53^{+0.07}_{-0.06}$ & $0.59^{+0.08}_{-0.08}$ & $0.64^{+0.07}_{-0.05}$\\
North & 13--26 & 1T &$0.43^{+0.10}_{-0.09}$ & $0.13^{+0.16}_{-0.13}$ & $0.38^{+0.06}_{-0.06}$ & $0.36^{+0.04}_{-0.04}$ & $0.50^{+0.06}_{-0.06}$ & $0.53^{+0.03}_{-0.03}$ \\
North & 13--26 & 2T & $0.63^{+0.14}_{-0.12}$ & $0.93^{+0.30}_{-0.27}$ & $0.61^{+0.12}_{-0.10}$ & $0.51^{+0.07}_{-0.06}$ & $0.62^{+0.08}_{-0.08}$ & $0.71^{+0.08}_{-0.06}$\\
North West
 & 7--28
&1T &
 $0.18^{+0.18}_{-0.17}$ & $0.55^{+0.30}_{-0.29}$ & $0.45^{+0.12}_{-0.11}$ & $0.40^{+0.07}_{-0.07}$ & $0.37^{+0.09}_{-0.09}$ & $0.49^{+0.04}_{-0.04}$ \\
North West 
 & 7--28
&2T &$0.33^{+0.25}_{-0.22}$ & $1.35^{+0.64}_{-0.49}$ & $0.70^{+0.21}_{-0.17}$ & $0.52^{+0.11}_{-0.09}$ & $0.44^{+0.13}_{-0.11}$ & $0.72^{+0.11}_{-0.10}$ \\
South 
 & 10--31
&1T&
$0.16^{+0.19}_{-0.16}$ & $0.00^{+0.18}_{-0.00}$ & $0.19^{+0.11}_{-0.11}$ & $0.23^{+0.07}_{-0.07}$ & $0.36^{+0.11}_{-0.12}$ & $0.34^{+0.04}_{-0.04}$ \\
South
 & 10--31
&2T&
 $0.28^{+0.27}_{-0.23}$ & $0.53^{+0.62}_{-0.49}$ & $0.40^{+0.22}_{-0.18}$ & $0.33^{+0.13}_{-0.11}$ & $0.45^{+0.16}_{-0.15}$ & $0.50^{+0.13}_{-0.11}$ \\
North East 
 & 17--36
&1T&
$0.40^{+0.17}_{-0.15}$ & $0.25^{+0.29}_{-0.25}$ & $0.42^{+0.11}_{-0.10}$ & $0.35^{+0.07}_{-0.07}$ & $0.43^{+0.12}_{-0.11}$ & $0.52^{+0.06}_{-0.05}$   \\
North East 
 & 17--36
&2T&$0.50^{+0.21}_{-0.18}$ & $0.66^{+0.52}_{-0.41}$ & $0.57^{+0.21}_{-0.15}$ & $0.44^{+0.13}_{-0.09}$ & $0.50^{+0.15}_{-0.14}$ & $0.64^{+0.15}_{-0.09}$   \\
Far North 
 & 20--42
&1T&
 $0.16^{+0.17}_{-0.15}$ & $0.00^{+0.20}_{-0.00}$ & $0.26^{+0.11}_{-0.10}$ & $0.24^{+0.08}_{-0.07}$ & $0.36^{+0.13}_{-0.13}$ & $0.42^{+0.05}_{-0.04}$   \\
Far North 
 & 20--42
&2T&$0.25^{+0.26}_{-0.18}$ & $0.46^{+0.60}_{-0.46}$ & $0.47^{+0.25}_{-0.17}$ & $0.35^{+0.15}_{-0.10}$ & $0.45^{+0.17}_{-0.16}$ & $0.61^{+0.15}_{-0.11}$ \\
\hline
\multicolumn{8}{l}{{\parbox{140truemm}{\footnotesize 
\par\noindent
\footnotemark[$\ast$]
F-test probability for adding second temperature component
\par\noindent
}\hss}}
\end{tabular}
\end{center}
\end{table*}

\subsection{XMM analysis}

\begin{figure}
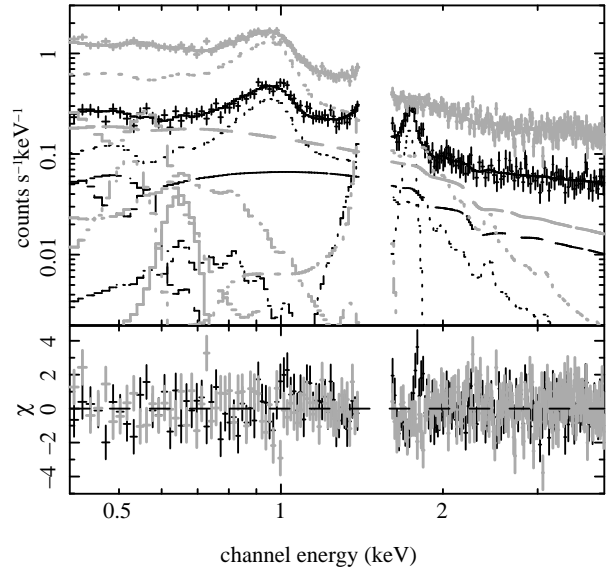

  \begin{center}
    \FigureFile(80mm,64mm){PASJ4026_figure3.ps}
  \end{center}
\caption{
Representative spectra of a square region  of MOS1 (black)
and PN (gray), fitted with the 1T ICM model.
 Dotted and other lines corresponds to the ICM components
and background, respectively.
}
\label{fig:xmm_spec}
\end{figure}

We used the spectra from MOS1, MOS2 and PN in the energy ranges of 0.4--4.0 keV,
but ignored the energy range of 1.4--1.6 keV.
The spectra in each region (pie and square)
 were fitted simultaneously with the 1T model for the ICM,
the Galactic emission, CXB,  and NXB components.
Here, the temperatures and normalizations of the Galactic components were
fixed at the best-fit values derived from the spectral fitting of the Suzaku data.
The abundance ratios were also fixed at the weighted averages of the four
offset fields observed with Suzaku (Section 4.3).
We modeled the NXB spectrum  with a powerlaw/b model
and two Gaussians for the  instrumental lines  at  1.48 keV (Al) and 1.74 keV (Si).
The powerlaw/b model is not folded through the ARF, and differs from a power-law.
We added two Gaussians at 0.56 keV and 0.65 keV for the solar wind charge exchange.
Figure \ref{fig:xmm_spec} shows representative spectra fitted in this manner.
We also fitted the spectra of deep fields in the same energy range
and confirmed that this background model reproduced the data well.
Most of  the spectra were fitted with the 1T model for the ICM with reduced $\chi^2\sim 1$.
We also applied the 2T model for the ICM.
Within 0.05 $r_{180}$, the 2T model showed a significantly decreased  $\chi^2$.
We adopted the Fe abundance from the 2T model fits within $0.05r_{180}$.

\section{Results}
\label{sec:res}

\subsection{Temperature distribution of the ICM}
\label{sec:res_kt}

Figure \ref{fig:fornax_kt} summarizes the temperatures of the
 $>6'$ regions of Suzaku and the pie regions of XMM versus radius of
NGC 1399, in comparison with those  from the central 6$'$ of NGC 1399 
by  \citet{Matsushita2007a}.
The temperatures derived from XMM and Suzaku are mostly consistent with each other.

The observed ICM temperatures are $\sim$ 1.2--1.3 keV at 0.03--0.1 $r_{180}$ 
which further decreases with radius to $\sim 1$ keV at 0.2--0.3 $r_{180}$.
The azimuthal dependence of the temperature is  largest at 0.1-0.2 $r_{180}$:
the ICM temperatures in the western and southern regions are 1.2$\sim$ 1.3 keV
and  those in the northern and eastern regions are $\sim$ 1 keV.
From the 2T model for the ICM, the temperatures of the two components
are 0.8--1 keV and 1.3--1.5 keV (Table \ref{tab:fornax_ktno}).

\begin{figure*}
\begin{center}
\centerline{
    \FigureFile(80mm,60mm){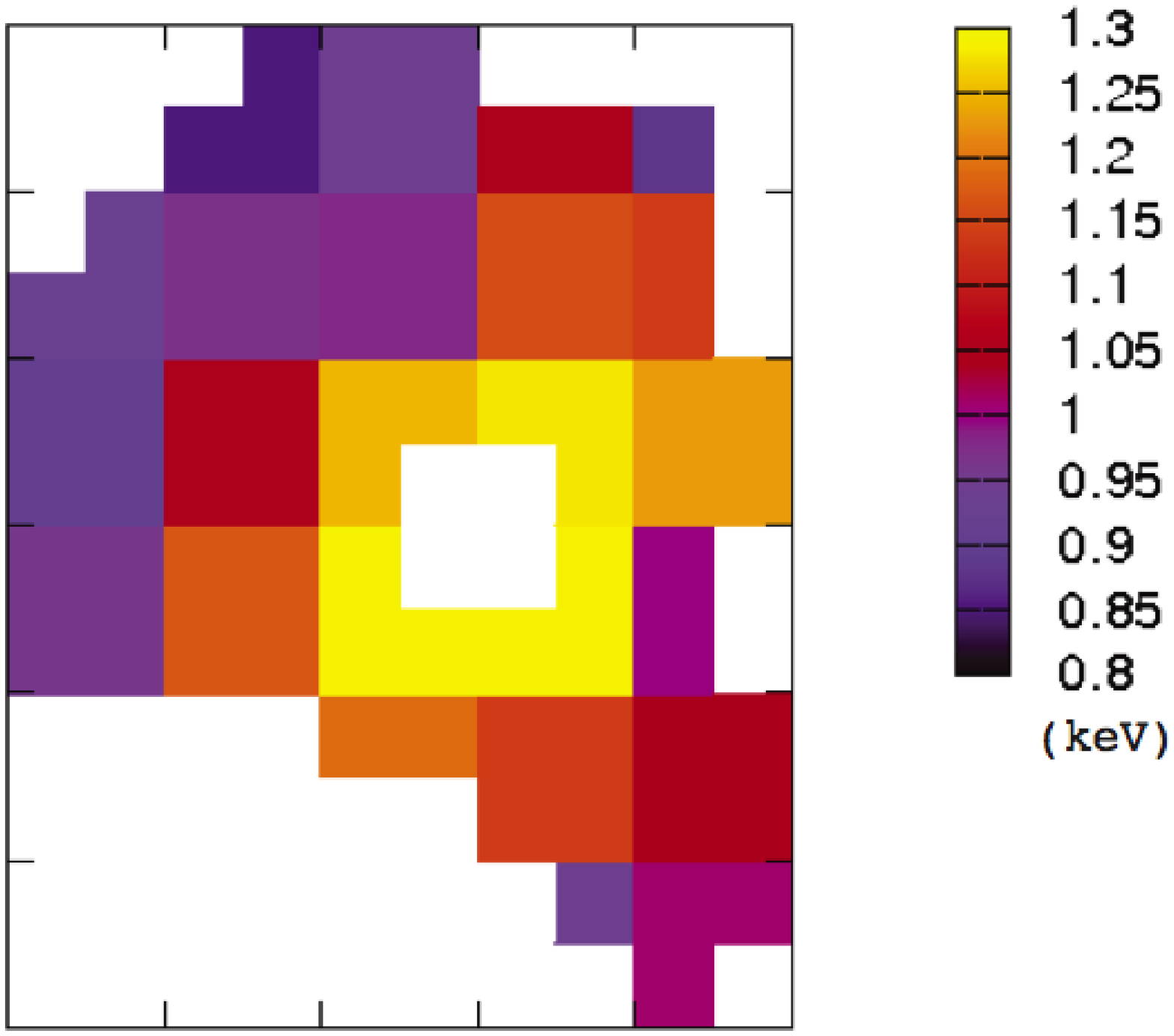}
\hspace{0.3cm}
    \FigureFile(75mm,60mm){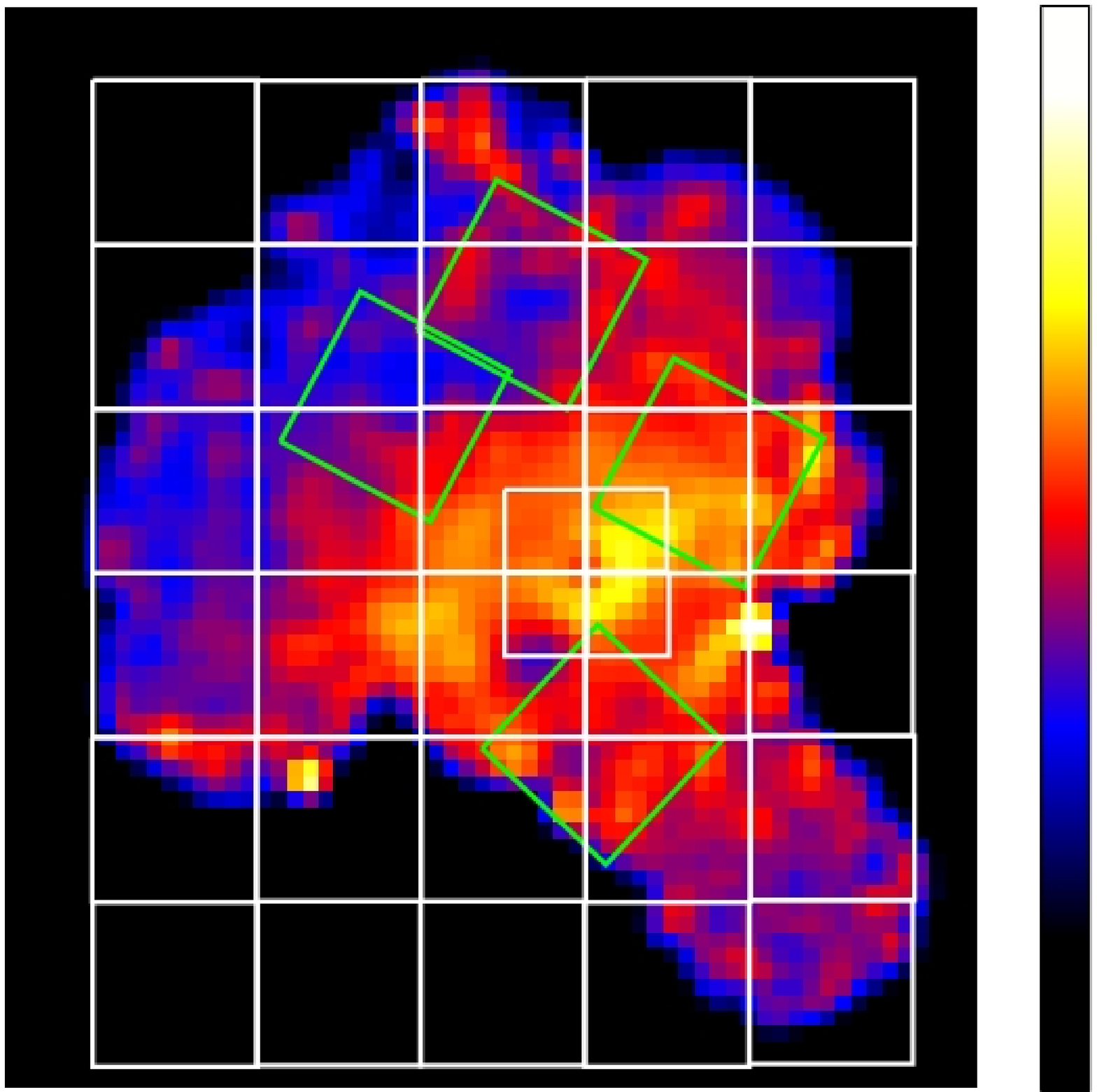}
}
\end{center}
\caption{
(left) Temperature map of the ICM derived from spectral fitting of the XMM data of the
square regions of 0.1$r_{180}\times$0.1$r_{180}$ (17.5$'\times$17.5$'$).
Temperatures of the four small square regions of 0.05$r_{180}\times$0.05$r_{180}$ surrounding NGC 1399 are not plotted,
because the 2T model  is required to fit the spectra.
(right)  Hardness ratio map of 1.0--1.2 keV to 0.7--1.0 keV.
White squares indicate the regions for the spectral analysis in the left panel.
Green squares show the four offset regions (South, North West, North East, Far North)
observed with Suzaku.
}
\label{fig:fornax_ktmap}
\end{figure*}

To study the temperature structure in more detail, 
the ICM temperature map derived from the square regions in the XMM 
observations
is shown in the  Figure \ref{fig:fornax_ktmap}.
Here, we did not plot the temperatures of four small square regions around the cD galaxy
because we need two temperatures to fit the spectra.
At a given distance from NGC 1399,
the southwestern regions show higher ICM temperatures than the
 northeastern regions.
We also derived a hardness ratio map of 1.0--1.2 keV to 0.7--1.0 keV observed
with XMM (Figure \ref{fig:fornax_ktmap}). The map  shows similar ICM temperatures.
The southwest and northwest regions tend to have higher ICM temperatures 
than the northeast regions.

\begin{figure}
\centerline{    \FigureFile(80mm,60mm){PASJ4026_figure4.ps}}
 \caption{The ICM temperatures  from Suzaku data
versus  radius of NGC 1399 in units of  $r_{180}$
for  the Center (black open circles),  
North (orange open circles),  Far north (red open circle), 
North East (green open circle), South (blue open circle), 
and North West (magenta open circle) fields derived from the 1T model fit for the ICM,
and those  from   the 2T  (black filled triangles) model fits within the central two radial bins.
The data of the innermost three  radial bins are
 from \citet{Matsushita2007a}.
The ICM temperatures of the pie regions of XMM  with the 1T model fits are plotted as diamonds. 
Red, green, blue and magenta correspond to the northern, eastern, southern and western pie regions, respectively.
}
\label{fig:fornax_kt}
\end{figure}

\subsection{Fe abundance distribution of the ICM}

The radial profiles of the Fe abundances observed with Suzaku and 
the pie regions of XMM derived from the 1T model fits are 
summarized  in Figure \ref{fig:fornax_fer}.
The values derived from the XMM and Suzaku data
are mostly consistent with each other,
although the  error bars for  Suzaku are smaller than those for XMM.

\begin{figure}
\begin{center}
    \FigureFile(80mm,60mm){PASJ4026_figure6.ps}
\end{center}
\caption{
Radial profile of the Fe abundance of the 
ICM derived from the 1T model fits for the
ICM using Suzaku data (open circles) and XMM data of the
pie regions (diamonds).
 Colors have  the same meanings as in Figure \ref{fig:fornax_kt}.
Data of the innermost three radial bins are from 
  \citet{Matsushita2007a}. Here, those
of  the innermost two radial bins are derived from the 2T model fits.
Region at 0.035--0.07 $r_{180}$ of the Center field  covers
mostly the south  of NGC 1399.
}
\label{fig:fornax_fer}
\end{figure}

\begin{figure}
\begin{center}
   \FigureFile(60mm,60mm){PASJ4026_figure7a_1007.ps}
   \FigureFile(60mm,60mm){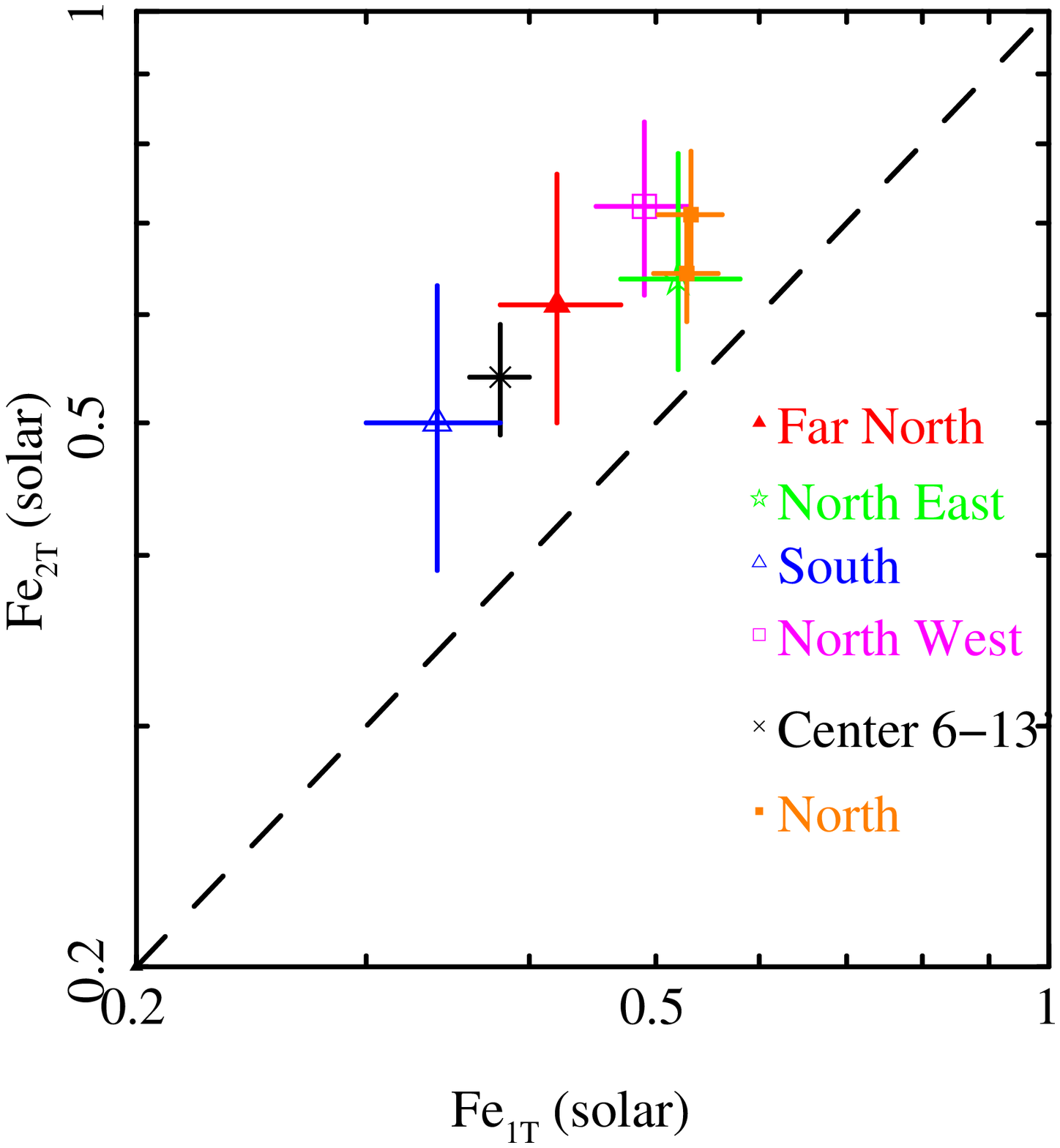}
\end{center}
\caption{
(upper panel)
Fe abundance of the ICM  observed with Suzaku
derived from the 2T model fits plotted against those from the 1T model fits.
Black cross, orange filled squares, blue open triangle, magenta open square, green star, and red filled triangle  correspond to
the Center, North, South, Northwest, Northeast, and Far North fields,
respectively.
(lower panel)
Radial profiles of the Fe abundance of the ICM derived from the 2T model fits of
the Suzaku data.
}
\label{fig:fornax_fe1t2t}
\end{figure}

In the  Suzaku data, at 0.03--0.2 $r_{180}$,
 the Fe abundances 
 of the North, North West, and North East fields derived from the 1T model for the ICM are about 0.5--0.6 solar,
which is  higher than the 0.3--0.4 solar value  of the South field.
The 2T model for the ICM gives  Fe abundances that 
are systematically higher  by $\sim$20--50\% than the 1T model
(Figure \ref{fig:fornax_fe1t2t}).
Although the errors became larger with the 2T model fits, the Fe abundances of the North and North West fields
still tend to be higher than that of the South field (Figure \ref{fig:fornax_fe1t2t}).

\begin{figure*}

\begin{center}
    \FigureFile(80mm,60mm){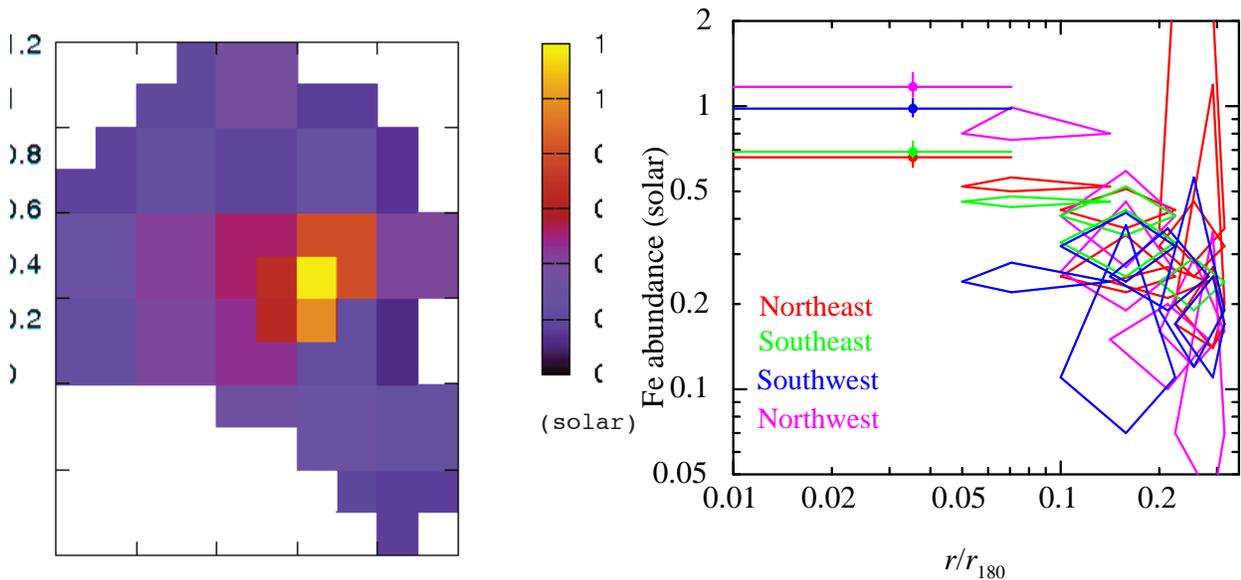}
\hspace{0.5cm}
    \FigureFile(80mm,60mm){PASJ4026_figure8b.ps}
\end{center}
\caption{
(left)
Fe abundance map of the ICM derived from the spectral fitting of the XMM data
of the square regions  
by the 2T model fits for the four central  small square regions of 0.05$r_{180}\times$0.05$r_{180}$, and
by the 1T model fits for the other regions of 0.1$r_{180}\times$0.1$r_{180}$.
NGC 1399 is located at the center of the four small square regions.
(right) Radial profile of the ICM of the square regions in the left panel
 derived from the 
2T model (filled circles with sold error bars) and 1T model (diamonds).
Red, green, blue and magenta colors correspond to the square regions
of northeast, southeast, southwest, and northwest, respectively, of NGC 1399.
}
\label{fig:fornax_femap}
\end{figure*}

With the 1T model fits of the XMM-Newton data, the Fe abundances of the ICM
 were derived out to 0.3 $r_{180}$.
As derived from the Suzaku observations,
at 0.05--0.2 $r_{180}$, the northern regions have higher Fe abundances of 0.5--0.6 solar,
whereas the south and west regions have lower values of 0.3--0.4 solar.
Beyond 0.2 $r_{180}$, the Fe abundance with the 1T model decreases to  $\sim$0.2--0.3 solar.
The left panel of Figure \ref{fig:fornax_femap} shows the Fe abundance map derived from the XMM data of the square regions,
and the right panel of Figure \ref{fig:fornax_femap} shows these abundances plotted
against the radius from NGC 1399.
Within the four central square regions, the Fe abundances are derived from 
the 2T model fit for the ICM.
The  southwestern region at $\sim 0.1 r_{180}$ has a lower Fe abundance
 than the  other regions at the same radius from  NGC 1399.

\subsection{Abundances of O, Mg, Si and S}
\label{sec:res_abund}

\begin{figure*}
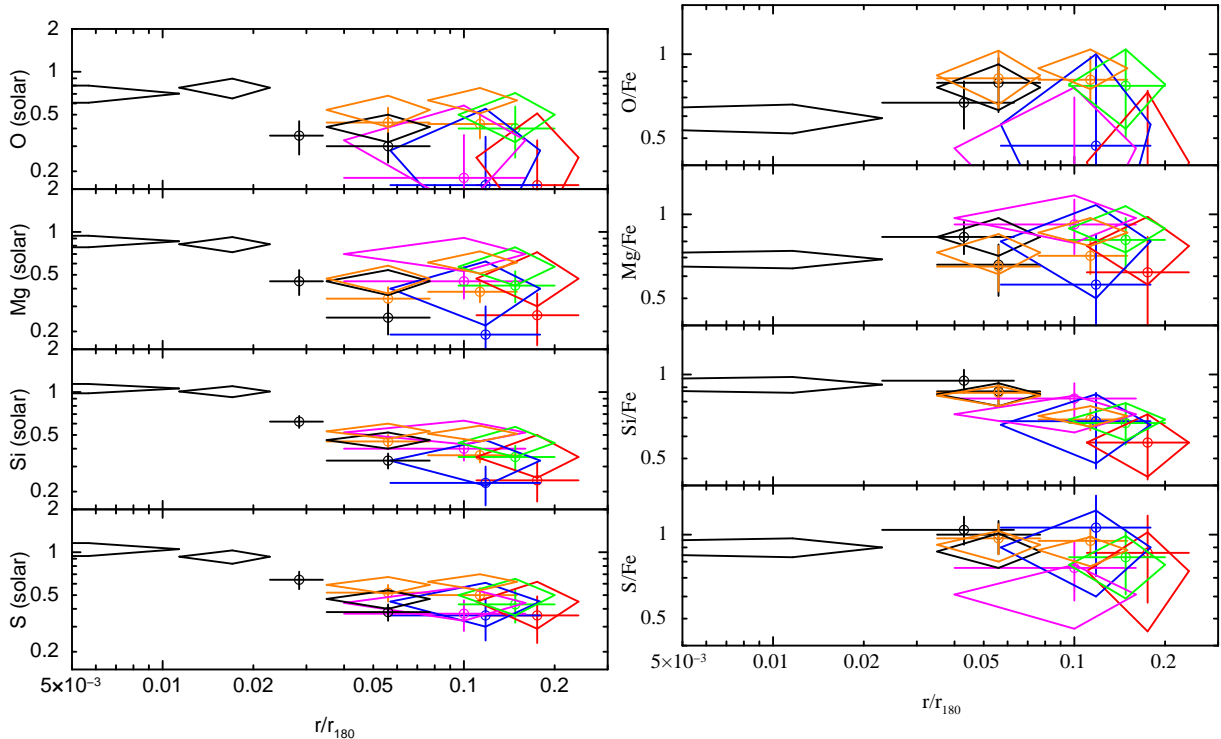

\centerline{
 \FigureFile(80mm,60mm){PASJ4026_figure9a.ps}
 \FigureFile(80mm,60mm){PASJ4026_figure9b.ps}
}
\caption{(left)
Abundance profiles of
  O,  Mg, Si, and S in the Center (black),
 North (orange),  North West (magenta), South (blue),
North East (green), and Far North (red) fields derived from the
1T (open circles)  or  2T (diamonds) model for the ICM.
Central three radial bins are derived by \citet{Matsushita2007a}.
(right) Radial profiles of the
  O/Fe,  Mg/Fe, Si/Fe, and S/Fe ratios in solar units.
The data within 0.02$r_{180}$ are weighted averages of the values of the
 two innermost radial bins.
}
\label{fig:fornax_omgsis}
\end{figure*}

Using the Suzaku observations, we also derived 
the abundances of O, Ne, Mg, Si, and S (Table \ref{tab:fornax_ktno}, Figure \ref{fig:fornax_omgsis}).
From 0.05 $r_{180}$ to 0.2 $r_{180}$,
the Mg, Si, and S abundances derived from the 1T model fits
are 0.2--0.5 solar, and 0.3--0.6 solar from the 2T model fits.
From the 1T model fits, the abundances of O are 0.2--0.4 solar, with fairly large
error bars, and the 2T model yields  O abundance values 
that are larger  by 0.1 solar.
When the 1T model is used, the derived Ne abundance values tend to be lower than those of other elements,
whereas with the 2T model, the Ne abundances are also close to those of Fe.

\begin{table*}
\begin{center}
\caption{
Weighted averages of the abundance ratios in  solar units derived from 
the Suzaku observations
}
\label{tab:fornax_aveabund}
\begin{tabular}{lcccccc}
\hline              
region & model & O/Fe &Ne/Fe & Mg/Fe & Si/Fe & S/Fe \\ \hline
0$'$--4$'$\footnotemark[$\ast$]  &  2T &
 $0.59^{+0.07}_{-0.07}$ &  -- &
 $0.69^{+0.05}_{-0.05}$ & 
 $0.92^{+0.06}_{-0.06}$ & 
 $0.90^{+0.07}_{-0.07}$ \\
6$'$--42$'$\footnotemark[$\dagger$]  &1T &
 $0.74^{+0.09}_{-0.09}$ & 
 $0.34^{+0.16}_{-0.17}$ & 
 $0.70^{+0.06}_{-0.05}$ & 
 $0.77^{+0.03}_{-0.03}$ &
 $0.94^{+0.06}_{-0.06}$ \\
6$'$--42$'$\footnotemark[$\dagger$]  &2T &
 $0.76^{+0.08}_{-0.08}$ & 
 $1.23^{+0.20}_{-0.21}$ & 
 $0.83^{+0.06}_{-0.06}$ & 
 $0.75^{+0.04}_{-0.03}$ &
 $0.84^{+0.06}_{-0.06}$ \\
6$'$--13$'$\footnotemark[$\dagger$]  &1T &
 $0.80^{+0.13}_{-0.13}$ & 
 $0.42^{+0.27}_{-0.27}$ & 
 $0.66^{+0.08}_{-0.09}$ & 
 $0.87^{+0.05}_{-0.06}$ &
 $0.99^{+0.08}_{-0.08}$ \\
6$'$--13$'$\footnotemark[$\dagger$]  &2T &
 $0.79^{+0.12}_{-0.11}$ & 
 $1.31^{+0.30}_{-0.30}$ & 
 $0.77^{+0.09}_{-0.09}$ & 
 $0.84^{+0.05}_{-0.05}$ &
 $0.90^{+0.09}_{-0.08}$ \\
7$'$--31$'$\footnotemark[$\dagger$]  &1T &
 $0.70^{+0.14}_{-0.13}$ & 
 $0.32^{+0.27}_{-0.25}$ & 
 $0.74^{+0.09}_{-0.09}$ & 
 $0.72^{+0.05}_{-0.06}$ &
 $0.91^{+0.09}_{-0.09}$ \\
7$'$--31$'$\footnotemark[$\dagger$]  &2T &
 $0.78^{+0.13}_{-0.13}$ & 
 $1.28^{+0.33}_{-0.33}$ & 
 $0.88^{+0.09}_{-0.09}$ & 
 $0.71^{+0.05}_{-0.05}$ &
 $0.82^{+0.08}_{-0.10}$ \\
17$'$--42$'$\footnotemark[$\dagger$]  &1T &
 $0.63^{+0.22}_{-0.22}$ & 
 $0.20^{+0.37}_{-0.20}$ & 
 $0.74^{+0.13}_{-0.13}$ & 
 $0.64^{+0.11}_{-0.11}$ &
 $0.83^{+0.21}_{-0.22}$ \\
17$'$--42$'$\footnotemark[$\dagger$]  &2T &
 $0.63^{+0.20}_{-0.18}$ & 
 $0.75^{+0.75}_{-0.75}$ & 
 $0.84^{+0.14}_{-0.13}$ & 
 $0.65^{+0.08}_{-0.08}$ &
 $0.77^{+0.17}_{-0.16}$ \\

\hline
\multicolumn{7}{l}{{\parbox{120truemm}{\footnotesize 
\par\noindent
\footnotemark[$\ast$]
weight average within 4$'$ from NGC 1399 from \citet{Matsushita2007a}
\par\noindent
\footnotemark[$\dagger$]
Weighted averages of abundance ratios at 6$'$--42$'$,  6$'$--13$'$ (Center and North),
\\7$'$--31$'$ (South and North West), and 17$'$--42$'$ (North East and Far North)}\hss}}
\end{tabular}
\end{center}
\end{table*}

Right panel of Figure \ref{fig:fornax_omgsis}
summarizes the abundance ratios of $\alpha$-elements divided by
the Fe value in solar units.
Because the abundances of $\alpha$-elements and Fe
are  correlated,
the errors of  the abundance ratios were estimated from 
\color{black}
confidence contours of each $\alpha$-element against Fe.
\color{black}
The 1T and 2T model fits give similar values for the O/Fe, Mg/Fe, Si/Fe,
 and S/Fe ratios.
The derived abundance ratios are mostly consistent with
the absence of radial gradients and azimuthal dependence,
 although the Si/Fe ratio shows a small negative radial gradient.
The weighted averages of the abundance ratios are calculated for the data
with similar radial ranges and are 
summarized in Table \ref{tab:fornax_aveabund}. 
The abundance ratios are mostly consistent with the absence of radial dependence.
When the
values beyond 6$'$ from the 1T model fits are averaged,
the abundance ratios of O/Fe, Mg/Fe, Si/Fe, and S/Fe 
are about 0.7, 0.7, 0.8, and 0.9  in solar units, respectively,
The 2T model yields similar abundance ratios within 10--15\%.

The weighted averages of Ne/Fe ratios from the 1T and 2T model fits differs by a factor of 3--4.
This is 
because the K-shell lines of Ne are completely mixed with the Fe-L lines,
and the Fe-L lines between the 1T and 2T models are different, which may
cause a discrepancy.
Therefore, 
the derived abundances of Ne might include fairly large systematic uncertainties.

\subsection{Surface brightness and gas density profiles}\label{brightness}

\begin{figure*}
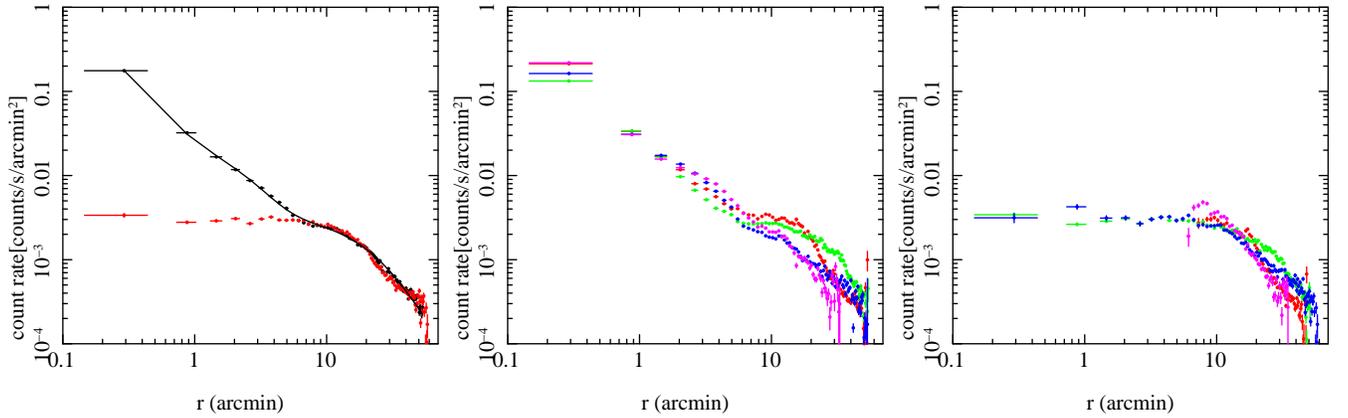

\centerline{
 \FigureFile(58mm,60mm){PASJ4026_figure10a.ps}
 \FigureFile(58mm,60mm){PASJ4026_figure10b.ps}
 \FigureFile(58mm,60mm){PASJ4026_figure10c.ps}
}
\caption{
(Left) Radial profiles of the surface brightness 
centered on NGC 1399 (black)  and on the cluster center (red) by
\citet{Pao2002}, which is $\sim$6$'$ northeast of NGC 1399.
The solid line represents the best-fit triple $\beta$-model.
(middle) Radial profiles of
 north (red), west (magenta), east (green), and south
(blue) sectors centered on NGC 1399.
(right) The same as the middle panel, but centered
on the cluster center.
}
\label{fig:radial}
\end{figure*}

We derived radial profiles of the X-ray surface brightness (left panel of 
Figure  \ref{fig:radial}) by azimuthally averaging the background-subtracted
intensity in the 0.8--1.2 keV energy band of MOS centered on NGC 1399, and on
the cluster center
derived by \citet{Pao2002}, which is $\sim$6$'$ northeast of NGC 1399.
Here, luminous point sources and a region around NGC 1404 were excluded.
For the profile centered on the cluster center, 
a region around NGC 1399 was also excluded.
Beyond 10$'$, the two radial profiles resemble  each other,
whereas within 10$'$,  a sharp brightness peak centered on NGC 1399 appears.
Similar to \citet{Pao2002}, we fitted the radial profile centered on NGC 1399
with a sum of three $\beta$-models that are all centered on NGC 1399.
As shown in Figure \ref{fig:radial}, the radial profile was roughly reproduced
with these three $\beta$-models.
Considering the similarity of the brightness profiles centered on NGC 1399
and on the cluster center beyond 10$'$, the density profile of the ICM from the
cluster center should  also be similar to that from NGC 1399.

To study the azimuthal variation in the radial brightness profile,
we divided the images centered on NGC 1399,
into four sectors--north, west, south, and east--
and derived radial brightness profiles within each sector.
The radial profiles of the four sectors are plotted in the middle panel
of Figure \ref{fig:radial}.  
At $\sim 20'$ ($\sim 0.1r_{180}$),  the brightness levels
in the north and east sectors
are higher by a factor of 2--3 than those in the south and west sectors.
We also derived radial brightness profiles in the four sectors
centered on the cluster center.
The discrepancy in the brightness levels  of the four sectors became 
smaller (right panel of Figure \ref{fig:radial}).

\section{Discussion}

\subsection{Comparison of the Fe abundance profiles with those of other systems}

\begin{figure}
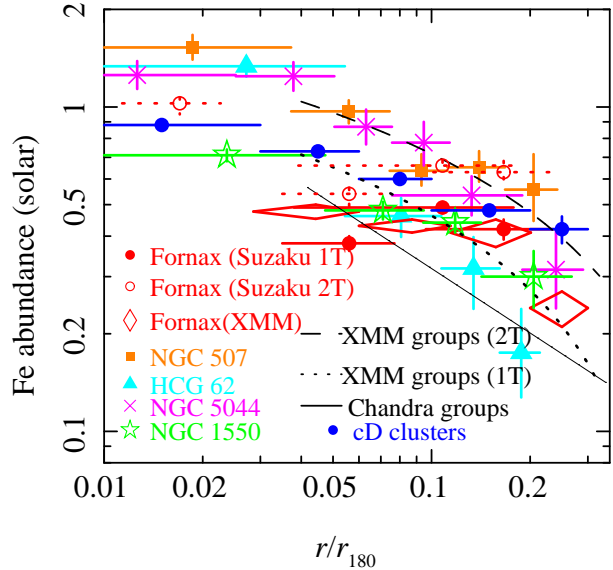

\begin{center}
 \FigureFile(80mm,60mm){PASJ4026_figure11.ps}
\end{center}
\caption{
Radial profiles of the Fe abundance in the Fornax cluster 
observed with Suzaku  derived from the
1T model (red filled circles with solid lines) and the 2T model (red open circles 
with dotted lines)
and with XMM from the 1T model (red diamonds).
Here, the Fe abundances with similar radial ranges are averaged.
Those of the NGC~507 group (orange filled squares; \cite{kSato2009a}),
HCG~62 group (cyan filled triangles; \cite{Tokoi2008}),
NGC~5044 group (magenta crosses; \cite{Komiyama2009}),
and a fossil group, NGC 1550 (green stars; \cite{kSato2010}).
The weighted average of relaxed clusters with a cD galaxy at their center
 observed with XMM (blue filled circles;\cite{Matsushita11}) and \color{black}
the best-fit regression relations for \color{black} groups of galaxies observed with Chandra
(black solid line: \cite{Rasmussen2007})  and cool-core groups observed
with  XMM (2T:black dashed line,1T:black
dotted line; \cite{Johnson11}) are also plotted.
}
\label{fig:fe}
\end{figure}

In Figure \ref{fig:fe},
the radial profiles of the
 Fe abundance in the Fornax cluster derived from the 1T and
2T model fits are
compared with those in other groups of galaxies observed with Suzaku:
the NGC 507 group \citep{kSato2009a},  HGC 62 group \citep{Tokoi2008},
the NGC 5044 group \citep{Komiyama2009},
 and the fossil group NGC 1550 \citep{kSato2010}.
Here, the Fe abundances of the Fornax cluster at similar radial ranges from NGC 1399
are averaged.
Considering the possible uncertainties in the Fe-L atomic data and background
components,
we cannot conclude that the 2T model fit is better than that the 1T model,
because the two models yield similar $\chi^2$ values in  the region with low surface brightness.

The  Fe abundance within 0.03 $r_{180}$ from NGC 1399
 is about 1 solar.
At 0.05 $r_{180}$ of the Fornax cluster,
the 1T and 2T model fits give the Fe abundances of
$\sim 0.4$ and $\sim 0.5$ solar, respectively.
These values are significantly smaller than the 
Fe abundances of NGC 5044 and NGC 507 groups,
at $\sim$0.05 $r_{180}$ which are derived from the 2T model fits.
At 0.1--0.2 $r_{180}$, the Fe abundances of the Fornax cluster are 
0.4 solar  and 0.6 solar, from the 1T and 2T model fits, respectively.
The value from the 2T model fits is close to those of the NGC 507 and NGC 5044 groups,
although HCG 62, a compact group of galaxies,  and NGC 1550, a fossil group,
tend to have smaller  Fe abundances when the 2T model fits are used.
These values are also similar to those of the clusters of galaxies 
(\cite{molendi2008}; \cite{Matsushita11}).
There is no systematic difference  between the
poor systems and the clusters of galaxies regarding
the Fe abundance of the ICM at 0.1--0.2 $r_{180}$

\color{black}
In Figure \ref{fig:fe}, the  Fe abundances  derived from the Suzaku data are compared with
those of the best-fit regression relations
 from Chandra data \citep{Rasmussen2007}
and from XMM data \citep{Johnson11}.
Here, we rescaled for the
differences in the definition of the solar abundance
table and virial radius.
\citet{Johnson11}  showed that
within $\sim$0.3 $r_{180}$, the 2T fit gives  abundances that are 
higher by a factor of $\sim$1.5  than those of the 1T fit.
Using the same temperature modeling of the ICM,
the Suzaku results agree very well with 
the regression relations from XMM data (\cite{Johnson11}).
In contrast, the results of Suzaku and XMM using the 2T model
are systematically higher than the best-fit regression line
of the groups derived with the Chandra data \citep{Rasmussen2007, Rasmussen2009},
where 2T model results are adopted
when the 2T model gave different abundances from the 1T model.
The Fe abundances by \citet{Rasmussen2007} show a significant scatter
at given radius in units of $r_{500}$ and some
 groups have comparable abundance profiles
 with those observed with Suzaku.
The radial profile of Fe abundance of the HCG 62 system with Suzaku
 is consistent with those of the Chandra groups.
Therefore, a major part of the differences in the Fe abundance
should be  caused by the differences in the sample.
Different  assumptions regarding the abundance ratio, i.e. VAPEC, or APEC, 
may not be responsible for  the discrepancy, since
the abundance patterns of these groups of galaxies 
derived with Suzaku do not different greatly from the solar ratio.
On the other hand, as shown in Komiyama et al. (2009), 
in regions with low surface brightness, 
uncertainties in the Galactic emission can also affect
the derived  Fe abundance.

\color{black}
\subsection{ICM abundance pattern  and contributions from SN
  Ia and SN II}
\label{sec:dis_abund}

Figure \ref{fig:atomnum}  summarizes the radial profiles of the O/Fe, Mg/Fe,
Si/Fe, and S/Fe ratios in five poor systems (the Fornax cluster, 
the NGC 5044 group, the HCG 62 group, and the NGC 507 group,
and a fossil group, NGC 1550), and clusters of galaxies, A262 ($kT\sim$2keV),
and AWM 7 ($kT\sim 4$keV).
The radial profiles
 of the Fornax cluster were the weighted averages of similar radial ranges
derived from the Suzaku data summarized in Table \ref{tab:fornax_aveabund}.
In particular, outside the cool cores at 0.05--0.2 $r_{180}$,
the scatter in the abundance ratios, except for the Mg/Fe ratios,
 among these systems is relatively small.
The scatter in the Mg/Fe ratios may be due to the systematic uncertainties 
in the Fe-L lines around the Mg-K lines.

In these poor systems, the abundance ratios exhibit
 no significant radial dependence.
The abundance pattern of O/Mg/Si/S/Fe in the offset regions of
the Fornax cluster is similar to those of the other groups of galaxies.
Out to 0.3 $r_{180}$
there is no hint of increase in the ratio of $\alpha$-elements to Fe abundances
in these poor systems.
These results indicate that both SN Ia and SN II products
have been mixed into the ICM, and that the ratio of the two types of SN
in the Fornax cluster is similar to those of the other groups of
clusters of galaxies.

\begin{figure}
\begin{center}
 \FigureFile(80mm,60mm){PASJ4026_figure12.ps}
\end{center}
\caption{
Radial profiles of O/Fe, Mg/Fe, Si/Fe, and S/Fe ratios
of the Fornax cluster (red)
from the 1T (diamonds) and 2T (filled circles) model fits.
Here, these values are weighted averages of similar radial ranges.
Radial profiles of 
the fossil group NGC 1550 (green; \cite{kSato2010}), the NGC 5044 group
(magenta;\cite{Komiyama2009}), HCG 62 (light blue; \cite{Tokoi2008}),
and the NGC 507 group (orange; \cite{kSato2009a}), and those
of the clusters of galaxies, AWM 7 (black; \cite{kSato2008}), and A262 (blue;\cite{kSato2009b}).
}
\label{fig:atomnum}
\end{figure}

\subsection{Radial profiles of the metal-mass-to-light ratios}
\label{sec:dis_mlr}

Because most of metals in the ICM synthesized in galaxies, the
metal-mass-to-light ratio is a useful measure for studying the chemical
evolution of clusters of galaxies.

To estimate the metal mass, we used the radial brightness profile derived 
in section \ref{brightness}, and derived the gas density and gas mass profiles.
Then, integrated  mass profiles of O  and Fe were
derived from the gas mass and abundance profiles using the
1T model fits. 
The 2T model fits yielded 10--20\% smaller normalizations of the ICM and
several tens of \% higher Fe abundances
in the Fornax cluster compared with  the 1T model fits.
Therefore, the total metal mass may have systematic uncertainties of 
several tens of percent, due to uncertainties in the temperature structure.

Because the K-band luminosity of a galaxy correlates well with the stellar
mass, we calculated the luminosity profile of the K-band.
We collected K-band magnitudes of galaxies  in 
$6\times 6~{\rm deg}^2$ box centered on NGC 1399 from the Two
Micron All Sky Survey (2MASS).
NGC 1399 has an apparent magnitude of 
$m_\mathrm{K} = 6.440$, or $\log{L_\mathrm{K}/L_\mathrm{K,\odot}} =
11.4$ using the 
foreground Galactic extinction $A_\mathrm{K} = 0.005$
 \citep{Schlegel1998} from the NASA/IPAC Extragalactic Database.
The average surface brightness in the region at
 $150' < r < 176'$ (0.86$r_{180} < r < 1.0r_{180}$) is subtracted as the background.
Within 0.3 $r_{180}$ of the Fornax cluster,
$\sim $10 early-type galaxies dominate the K-band luminosity.
NGC 1399 dominates the luminosity in the region
$r<\sim 0.04~r_{180}$.
At $r\sim 0.06 r_{180}$, NGC 1404, a bright galaxy, causes
 a break in the luminosity profiles.
At $\sim$0.2 $r_{180}$, several early-type galaxies including 
NGC 1380 contribute to the luminosity profiles.

The integrated mass-to-light ratios for O and Fe
(OMLRs and IMLRs) using the K-band luminosities are
summarized in  Figure \ref{fig:mlr_radial_all}.
The error bars of the mass-to-light ratios include only
abundance errors.
The OMLR and IMLR profiles are not smooth because of several luminous galaxies
 in the Fornax cluster.
The profiles increase with radius out to $0.2r_{180}$.
However, the IMLR profile became flat from 0.2 $r_{180}$ to $0.3r_{180}$,
due to increase of K-band luminosity from several luminous early-type galaxies.
We extrapolated the best-fit three $\beta$-model of the surface brightness
of the ICM, \color{black} assuming that Fe abundance of the ICM beyond 0.3 $r_{180}$ is the same
as in that of the best-fit value at 0.2--0.3 $r_{180}$.
\color{black}
Then,  at 0.5 $r_{180}$, the IMLR may not increase very much (Figure \ref{fig:mlr_radial_all}).

Figure \ref{fig:mlr_radial_all} compares the 
derived OMLRs and IMLRs of the Fornax cluster
with those of the NGC 5044 group \citep{Komiyama2009}, 
the fossil group NGC 1550 \citep{kSato2010},  and the Abell 262 cluster 
($kT\sim 2$ keV; \cite{kSato2009b}), and IMLR of the AWM 7 cluster ($kT\sim 4$ keV;\cite{kSato2008}).
In contrast to the Fornax cluster, 
central galaxies dominates the K-band luminosity
in the NGC 5044 group, the fossil group NGC 1550, and Abell 262,
 and most of the other galaxies are dwarfs. 
Therefore, the profiles of these systems are smoother than those
of the Fornax cluster.
Within 0.1 $r_{180}$, the IMLR and OMLR of the Fornax cluster are
much smaller than those of the other systems.
The similar IMLR profiles of the other systems within 0.1 $r_{180}$
indicate that the recent metal supplies from the central galaxies are typical.
The small IMLR in the Fornax cluster indicates
that the accumulation time scale of metals from the cD galaxy should be shorter,
reflecting the fact that the cD galaxy is 
not located at the cluster center and moving.
At 0.1--0.3 $r_{180}$, the IMLR of the Fornax cluster is still an order of
magnitude smaller than those of the rich systems, AWM 7, Abell 262, 
and NGC 1550. The IMLR of these three systems increase with radius
in the same way from 0.1 $r_{180}$ to 0.3--0.5 $r_{180}$.
However, the IMLRs of the NGC 5044 group is constant from 0.1 $r_{180}$
to 0.3${r_{180}}$, and at 0.3 $r_{180}$, the IMLR of the NGC 5044 group
is nearly an order of magnitude larger than that of the Fornax cluster.
The extrapolated value of the IMLR at 0.5 $r_{180}$ of the Fornax cluster
are still over an order of magnitude smaller than those of the NGC 1550 group
and Abell 262 cluster.

\citet{Rasmussen2009} found that the IMLR derived from Chandra and Suzaku
are consistent with each other,  despite  the systematic difference
in the derived Fe abundance.
This is because a larger Fe abundance is associated with a smaller gas mass
in lower temperature groups.

\begin{figure*}
\begin{center}
 \FigureFile(80mm,60mm){PASJ4026_figure13a_20110928.ps}
 \FigureFile(80mm,60mm){PASJ4026_figure13b.ps}
\end{center}
\caption{
Radial profiles of integrated
 IMLR (left panel) and  OMLR (right panel) 
 in  the K-band of the Fornax cluster (red filled circles),
NGC 5044 group (purple open circle;\cite{Komiyama2009}), NGC 1550 group
(green crosses;\cite{kSato2010}),  
Abell 262 cluster (blue filled triangles;\cite{kSato2009b}), and AWM 7 cluster (black open squares;\cite{kSato2008}).
Dotted line for the Fornax cluster represent extrapolated values using the
surface brightness and the Fe abundances within 0.3 $r_{180}$
}
\label{fig:mlr_radial_all}
\end{figure*}

\subsection{Dynamical history of the Fornax cluster}

Chandra observations revealed that NGC 1399 is moving within the Fornax cluster \citep{Scharf2005}.
The X-ray emission of the northeast region of the Fornax cluster is
brighter than that of the south region.
The ICM temperature of the southwest region is higher
than that of the northeast region.
The lower ICM temperature and the higher brightness of the northeast region
indicate lower entropy in the ICM.
The Fe abundance of the northeast region at 0.1--0.2 $r_{180}$ is higher
 than that of the southwest region at the same radius.
Lower entropy and the higher Fe abundance are usually observed in the cool cores.
Therefore, 
the cD galaxy may have traveled from the center of the cluster to the south
due to   recent dynamical evolution, as suggested by
 the optical dynamical observations 
by \citet{Dunn2006}.

Recent dynamical evolution 
 might have  hindered the strong concentration of
hot gas in the central region of the Fornax cluster.
The X-ray luminosity within 4$r_e$ of  NGC 1399 is smaller
by  a factor of 20  than that of
NGC 5044 (\cite{Matsushita01}; \cite{Nagino2009}).
The X-ray luminosity of 
NGC 1550 is also higher by an order of magnitude  than  that of NGC 1399
(\cite{Fukazawa2006}).
For NGC 1399, the central Fe peak is narrower and the IMLR 
is much smaller than those of  the other groups.
Some merging clusters also have smaller scale of Fe peaks (\cite{Matsushita11}).
During cluster merging, 
mixing of the ICM could destroy the central Fe peak. 
The Fornax cluster is also in a stage of dynamical evolution,
and may be in  a phase of  central Fe peak destruction.
\color{black}
Then, recent supply of metals  from NGC 1399 via stellar mass loss and SN Ia
 produces a smaller Fe abundance peak 
than in  the other groups.
\color{black}

\subsection{Metal enrichment and feedback in the groups and clusters}
The metal distribution in the ICM can be a powerful tracer of the history of 
 gas heating in the early epoch, because the relative timing of metal enrichment and heating should affect the present amount and distribution of the 
metals in the ICM. 
The observed IMLRs in poor systems are scattered by an order of magnitude,
whereas, the Fe abundance profiles are similar among these systems and also
similar to those of  clusters.
 The Fornax cluster has the smallest IMLR out to 0.3 $r_{180}$.
If  all galaxies synthesize a similar amount of metals per unit stellar mass,
the observed low IMLR and similar Fe abundance compared with other systems indicate
that  a significant fraction of the Fe was synthesized in an early phase of  cluster evolution. 
 If metal enrichment occurred before  energy injection, the poor systems would carry relatively smaller metal mass with a smaller gas mass than  rich clusters, whereas, the metal abundance would be quite similar to those in rich clusters. 
Dynamical evolutions  may also  change the gas distribution, but  not 
the abundance of the gas.
In contrast, if metal enrichment occurred after  energy injection, the metal mass becomes comparable to those in rich clusters and indicates a higher abundance reflecting a
lower gas mass.

Similar to rich clusters,
most of the stellar light in the Fornax cluster originates from bright old early-type galaxies (\cite{Kunt2000}). The stellar metallicity and [Mg/Fe] ratios
of the central regions of these galaxies are similar to those of similar size in rich clusters (\cite{Kunt2000}).
The O/Mg/Fe abundance pattern of the hot insterstellar medium (ISM) of NGC 1399 and NGC 1404 in the Fornax cluster derived from the Suzaku observations
(\cite{Matsushita2007a})
are similar to those of NGC 4636 (\cite{Hayashi09}) in the Virgo cluster.
Reflection grating spectrometer (RGS)
 observations onboard the XMM showed that the O/Fe ratio of the ISM in  NGC 1404
in the Fornax cluster is
similar to those of elliptical galaxies in the Virgo cluster (\cite{Werner09}).
Because the hot ISM in elliptical galaxies results from the 
 accumulation of stellar mass loss
and the ejecta from recent SNe Ia, 
the observed similarity of the abundance patterns of
the hot ISM suggests that the stellar metallicity and SN Ia rate do
not  different greatly
between the present elliptical galaxies in the Fornax and Virgo clusters.
These results indicate that stars in  elliptical galaxies in the Fornax cluster
were enriched  in the same way as those in rich clusters.

The observed higher stellar mass fraction and the lower gas mass fraction
 within $r_{500}$ in poor systems,
are sometimes interpreted as demonstrating that the
 star formation efficiency depends on the system mass.
However, the similarity in abundances in the ICM between groups and clusters and
a scatter in the metal-mass-to-light ratios can be better explained by early metal
enrichment, as discussed in \citet{Matsushita11},
on the basis of the
 relatively flat abundance profiles
of Fe in the ICM  in clusters of galaxies.

\section{Summary and Conclusion}

Suzaku and the XMM observed the Fornax cluster out to 0.2--0.3 $r_{180}$,
and derived the temperature and  O, Mg, Si, S, and Fe abundances in the ICM
quite accurately. 
The ICM temperature decreases from 1.5 keV near the cD galaxy
NGC 1399  to 1 keV in the outer region.
 The Fe abundance around NGC 1399
 is about one solar, and the central Fe abundance peak is narrower than 
that in  the groups and clusters with cool cores.
The Fe abundance  drops to 0.3--0.5~solar at 0.1--0.2~ $r_{180}$, which is similar
to the values in  other groups and clusters.
 The abundance ratios, O/Fe,  Mg/Fe, Si/Fe,
 and S/Fe, are close to the solar ratio, similar to
other groups and clusters of galaxies.
The abundance pattern indicates that both SN Ia and SN II products
have been mixed into the ICM, and the ratio of the two types of SN
in the Fornax cluster is similar to those of other groups and
clusters of galaxies.

The northeast region of  NGC 1399 has a lower ICM temperature, higher 
abundances and higher surface brightness than the southwest regions.
Therefore, 
the cD galaxy may have traveled from the center of the cluster to the south
owing to  recent dynamical evolution.

Out to 0.3 $r_{180}$, the IMLR of the Fornax cluster is 
 an order of magnitude smaller than that of other groups
and clusters.
The metal distribution
in the ICM can be used as a tracer of the past history of heating and
enrichment in  clusters of galaxies.
Scatter in the IMLR and similarity in the abundances in the ICM indicate
early metal synthesis in groups and clusters.

\bigskip


\end{document}